\documentclass[aps,epsf,preprint,nofootinbib]{revtex4}

\newcommand{\nslash}{\kern 0.2 em n\kern -0.50em /}

\newcommand{\beq}{\begin{eqnarray}}
\newcommand{\eeq}{\end{eqnarray}}

\def\bq{\begin{eqnarray}}
\def\eq{\end{eqnarray}}
\def\bq{\begin{equation}}
\def\roughly#1{\mathrel{\raise.3ex\hbox{$#1$\kern-.75em
\lower1ex\hbox{$\sim$}}}}

\begin{document}

\preprint{\hfill\parbox[b]{0.3\hsize}
{ }}

\def\bra{\langle }
\def\ket{\rangle }

\title{A quark model analysis of the Sivers function}

\author{A. Courtoy$^1$, F. Fratini$^2$,
S. Scopetta$^{2,3}$\footnote{corresponding author:
E-mail address: sergio.scopetta@pg.infn.it}, V. Vento$^{1,4}$}
\affiliation
{\it
(1)
Departament de Fisica Te\`orica, Universitat de Val\`encia
\\
and Institut de Fisica Corpuscular, Consejo Superior de Investigaciones
Cient\'{\i}ficas
\\
46100 Burjassot
(Val\`encia), Spain
\\
(2)
Dipartimento di Fisica, Universit\`a degli Studi di Perugia, via A. Pascoli
06100 Perugia, Italy.
\\
(3) INFN, sezione di Perugia, via A. Pascoli
06100 Perugia, Italy
\\
(4) TH-Division, PH Department, CERN,
CH-1211 Gen\`eve 23, Switzerland
}

\begin{abstract}

We develop a formalism to evaluate the Sivers function. 
The approach is well suited for
calculations which use constituent quark models to describe the
structure of the nucleon. A non-relativistic reduction of the
scheme is performed and applied to the Isgur-Karl model of hadron
structure. The results obtained are consistent with a sizable Sivers
effect and the signs for the $u$ and $d$ flavor contributions turn
out to be opposite. This pattern is in agreement with the one found
analyzing, in the same model, the impact parameter dependent
generalized parton distributions.
The Burkardt Sum Rule turns out to be fulfilled to a large extent.
We estimate the QCD evolution of
our results from the momentum scale of the model to the experimental
one and obtain reasonable agreement with the available data.

\end{abstract}

\maketitle

\section{Introduction}

The partonic structure of transversely polarized nucleons
is one of their less known features
(for a review, see, e.g.,
Ref. \cite{bdr}).
Nevertheless, experiments for its determination are progressing very fast
and the relevant experimental
effort
has motivated a strong theoretical
activity (for recent developments, see Ref. \cite{olmo}).
The present work aims to contribute to this effort
by using a successful theoretical scenario for the calculation of the
Sivers function.

Semi-inclusive deep inelastic scattering (SIDIS), i.e.
the process $A(e,e'h)X$, with the
detection in the final state
of a produced hadron $h$ in coincidence
with the scattered electron $e'$, is one of the proposed
processes to access the parton distributions (PDs)
of transversely polarized hadrons.
For several years it has been known that
SIDIS off a transversely polarized target
shows azimuthal asymmetries,
the so called ``single spin asymmetries'' (SSAs) \cite{Collins}.
As a matter of fact, it is predicted that
the number of produced hadrons in a given direction or in the opposite
one, with respect to the reaction plane,
depends on the orientation of the transverse spin
of a polarized target with respect to the direction
of the unpolarized beam. It can be shown that the SSA
in SIDIS off transverse polarized targets is essentially
due to two different physical mechanisms,
whose contributions can be technically distinguished
\cite{mu-ta,ko-mu,boer,bac1}.
One of them is the Collins mechanism, due to parton final state interactions
in the production of a hadron
by a transversely polarized quark
\cite{Collins}, and will not be discussed here.
The other is the Sivers mechanism \cite{sivers},
producing
a term in the SSA which is given by the product of
the unpolarized fragmentation function with
the Sivers PD,
describing the number density of unpolarized quarks
in a transversely polarized target.
The Sivers function is
a Transverse Momentum Dependent (TMD) PD;
it is a time-reversal odd object \cite{bdr} and for this reason,
for several years, it was believed to vanish due to time reversal invariance.
However, this argument was invalidated by a calculation
in a spectator model \cite{brohs}, following the observation
of the existence of leading-twist
Final State Interactions (FSI) \cite{brodhoy}.
The current wisdom is that a non-vanishing Sivers function
is generated by the gauge link in the definition of TMD
parton distributions \cite{coll2,jiyu,bjy}, whose
contribution does not vanish in the light-cone gauge,
as  happens for the standard PD functions.
For the same reason it is difficult to relate
the Sivers Function to the target helicity-flip,
impact parameter dependent (IPD), generalized parton
distribution (GPD) E. Although simple relations between
the two quantities are found in models
\cite{burk,sch},
a clear model independent formal relation is still
to be proven, as shown in Ref. \cite{mgm}.

Recently, the first data
of SIDIS off transversely polarized targets have been published,
for the proton \cite{hermes} and the deuteron \cite{compass}.
It has been found that, while the Sivers effect
is sizable for the proton, it becomes negligible for the deuteron,
so that apparently the neutron contribution cancels the proton one,
showing a strong flavor dependence of the mechanism.
Experiments on transversely polarized
$^3$He target, aimed at extracting the neutron information, addressed
in \cite{brod}, are being performed at JLab \cite{06010,06011}.
A realistic calculation of nuclear effects for a proper
extraction of the neutron
information has also been performed \cite{scop}.
Different parameterizations of the available
SIDIS data have been published \cite{ans,coll3,Vogelsang:2005cs}, 
still with large
uncertainties.
Further analyses are in progress (see, i.e. \cite{bog}).
New data,
which will
reduce the uncertainties on the extracted Sivers function
and will help discriminate between different theoretical predictions,
will be available soon.

This experimental scenario motivates the formulation of theoretical
estimates. One would like to perform a calculation from first
principles in QCD, however this is not yet possible. Lacking this
possibility it becomes relevant to perform model calculations of the
Sivers function. Several estimates exist, in a quark-diquark model
\cite{brohs,jiyu,bacch}; in the MIT bag model, in its simplest
version \cite{yuan} and introducing an instanton contribution
\cite{d'a}; in a light-cone model \cite{luma}; in a nuclear
framework, relevant to establish the manifestation of the Sivers
function in proton-proton collisions \cite{bianc}.

To our knowledge, no calculations of the Sivers function have been
performed in a Constituent Quark Model (CQM), i.e. a model described
in terms of constituent quarks and whose properties have been fixed
from hadronic observables. The CQMs have a long history of
successful predictions in studies of the hadronic spectrum and the
low energy electroweak structure of hadrons. Ascribing a scale to
the model calculations \cite{pape,jaro} and using QCD evolution
\cite{dokshitzer,altarelli} one can evolve the leading twist
component of the observable calculated in this low energy scale to
the high momentum one where DIS experiments are carried out. Such
procedure has proven successful in describing the gross features of
PDs (see, e.g., \cite{trvv,h1,oam}) and GPDs (see, e.g.
\cite{epj,bpt}), by using different CQMs. Similar expectations
motivate the present study of the Sivers function.

In here we propose a formalism to calculate the valence quark
contribution to the Sivers function from any CQM. Thereafter, we
choose the Isgur-Karl model \cite{ik} to perform a detailed
calculation in order to describe the performance of the approach. A
difference in the calculation of TMDs, with respect to calculations
of PDs and GPDs, is that the leading twist contribution to the
one-gluon-exchange (OGE) FSI has to be evaluated. This is done
through a non-relativistic (NR) reduction of the relevant operator,
according to the philosophy of constituent quark models \cite{ruju}.

The paper is structured as follows. In the second section, the main
quantities of interest are introduced. In the following section, the
formalism for the calculation of the Sivers function in a CQM is
developed. The Isgur--Karl model is presented in the fourth section,
together with the numerical results of the calculation and their
discussion. The following section is devoted to the QCD evolution of
the model results and to the comparison with the available data. In
the last section we draw conclusions from our study.

\section{The theoretical framework}

The Sivers
function, $f_{1T}^{\perp {\cal Q} } (x, k_T)$,
the quantity of interest here,
is formally defined, 
according to the Trento convention
\cite{trento,goeke},
for the quark of flavor ${\cal Q}$,
through the following
expression\footnote{Here and in the
following, $a^\pm = (a_0 \pm a_3)/ \sqrt{2}$ and $k_T=|\vec{k}_T|$.}:
\begin{eqnarray}
\Phi^{\cal Q}(x,\vec k_T,S) & = &
f_1^{\cal Q}(x, k_T) -
{\epsilon_T^{ij}k_{Ti}S_{Tj} \over M}
f_{1T}^{\perp {\cal Q}} (x, k_T)
\nonumber
\\
& = & {1 \over 2} \int { d \xi^- d^2 \vec \xi_T
\over (2 \pi )^3}
e^{-i ( x \xi^- P^+  - \vec{\xi}_T \cdot {\vec k_T} )}
\langle P,S | \hat O_{\cal Q} | P, S \rangle
~,
\label{def0}
\end{eqnarray}
where
$ {\vec S_T} $ is the transverse spin of the target hadron,
the normalization of the covariant spin vector is $S^2 = -1$,
$M$ is the target mass and $f_1^{\cal Q}(x, k_T)$ 
is the $k_T-$dependent unpolarized
PD.
The operator $\hat O_{\cal Q}$ is defined as follows \cite{jiyu,bjy}:
\begin{eqnarray}
\hat O_{\cal Q} =
\bar \psi_{\cal Q}(0,\xi^-,\vec{\xi}_T)
{\cal L}^\dagger_{\vec{\xi}_T}(\infty,\xi^-)
\gamma^+
{\cal L}_0(\infty,0)
\psi_{\cal Q}(0,0,0)~,
\label{o_siv}
\end{eqnarray}
where $\psi_{\cal Q}(\xi)$ is the quark field and
the gauge link is:
\begin{eqnarray}
{\cal L}_{\vec{\xi}_T}(\infty,\xi^-) =
P \exp \left (-ig \int_{\xi^-}^\infty A^+(\eta^-,\vec{\xi}_T) d \eta^-
\right )~,
\label{link}
\end{eqnarray}
where $g$ is the strong coupling constant.
One should notice that this definition for the gauge link
holds in covariant (non singular) gauges,
and in SIDIS processes, since the definition of the
Sivers function is process dependent.
As observed in Ref. \cite{brohs} for the first time,
and later confirmed
using factorization theorems in \cite{Ji:2004wu,Collins:2004nx},
the gauge link, which represents the exchange of gluons,
provides a scaling contribution which  makes the Sivers function  non vanishing
in the Bjorken limit.

Taking the proton polarized along the $y$ axis
one has therefore:
\begin{eqnarray}
f_{1T}^{\perp {\cal Q}} (x, {k_T} ) =
-
{
M \over 4 k_x
}
\int
{ d \xi^- d^2 \vec {\xi}_T \over (2 \pi)^3 }
e^{-i ( x \xi^- P^+  - \vec{\xi}_T \cdot {\vec k_T} )}
\langle \hat O_{\cal Q} \rangle~,
\label{kt-siv}
\end{eqnarray}
where the following matrix element has been defined:
\begin{eqnarray}
\langle \hat O_{\cal Q} \rangle
%& = & 
=
\{ \,
\langle P S_y = 1 | 
\hat O_{\cal Q} 
| P S_y = 1 \rangle 
%\\
%& - &
- 
\langle P S_y = - 1 | 
\hat O_{\cal Q} 
| P S_y = - 1 \rangle \, \}~.
\label{okt-siv} 
\end{eqnarray}

Considering a helicity basis for the target, the
Sivers function Eq. (\ref{kt-siv}) can be written:
\begin{eqnarray}
f_{1T}^{\perp {\cal Q}} (x, {k_T} ) =
\Im \left \{
{
M \over 2 k_x
}
\int
{ d \xi^- d^2 \vec{\xi}_T \over (2 \pi)^3 }
e^{-i ( x \xi^- P^+  - \vec{\xi}_T \cdot {{\vec k_T}} )}
\langle P S_z = 1 |
\hat O_{\cal Q}
| P S_z = - 1 \rangle \right \}~.
\label{work}
\end{eqnarray}
This equation, finite in the limit
of $k_x \rightarrow 0$,
will be used to evaluate the Sivers function,
using a CQM to describe the proton.
We will now proceed to expand the gauge link,
Eq. (\ref{link}), in the coupling constant,
$g$ ~:
\begin{equation}
P \exp \left (-ig \int_{\xi^-}^\infty A^+(\eta^-,\vec{\xi}_T) d \eta^-
\right ) = 1 - ig \int_{\xi^-}^\infty A^+(\eta^-,\vec{\xi}_T) d \eta^-  + ...
\label{expa}
\end{equation}
 If  the gauge link were not taken into account, 
it is clear from Eqs.~(\ref{o_siv})-(\ref{work}) 
that the matrix element Eq. (\ref{okt-siv}) would be zero
and the Sivers function would vanish. 
For this reason, the first  term  on the right-hand side of
Eq.~(\ref{expa}) 
does not contribute to the Sivers function.

A few theoretical predictions have been formulated 
for the Sivers function. Let us recall two of them.

The first one,
based on rather general principles,
is the so called Burkardt Sum Rule 
\cite{Burkardt:2004ur}, 
stating that
the total average transverse momentum of the partons in a hadron,
$\langle \vec k_T \rangle$,
which
can be defined in terms of the sum of the first moments of
the Sivers function for all the partons in the target,
has to vanish.

The second one is the
conjecture according to which
the Sivers function
could be related to the formalism of the
IPD GPDs \cite{burk1}, although, as it has been
discussed in the Introduction,
simple relations between
the two quantities are found only in models
\cite{burk,sch} and
a clear model independent formal relation is still
to be proven \cite{mgm}.
The IPD GPDs are the
Fourier transform 
of the GPDs with respect to the
transverse momentum transfer $\vec \Delta_T$, at vanishing skewness $\xi$.
In the case of the helicity independent
GPD, $H_{\cal Q}(x,\xi,\Delta^2)$, one has:
\begin{equation}
H_{\cal Q}(x,\xi=0,b^2) = \int { d^2 \vec \Delta_T \over (2 \pi)^2 }
e^{-i {\vec b} \cdot \vec \Delta_T }
H_{\cal Q}(x,\xi=0,\Delta^2)~,
\label{ipd}
\end{equation}
and analogous definitions hold for the helicity independent,
target spin-flip GPD $E_{\cal Q}(x,\xi,\Delta^2)$ and for the other GPDs.
It has been shown that these quantities have a probabilistic interpretation,
describing the location of the quarks of flavor
${\cal Q}$ in the transverse plane
and providing us with a three dimensional picture of the proton \cite{burk1}.
In Ref. \cite{burk2,die}
(see also Ref. \cite{3d} for a recent review on this subject),
it has also been shown that, in a transversely polarized
proton, for example along the $y$ direction, the quantity describing
the distribution of the quarks of flavor ${\cal Q}$, with
longitudinal momentum $x$, in the transverse plane,
independently of their helicity, is
\begin{equation}
\tilde \rho_{\cal Q}(x,\xi=0,\vec b) = {1 \over 2} H_{\cal Q}(x,0,b^2)
- {b_x S_y \over 2 M} {d \over d b^2} E_{\cal Q}(x,0,b^2)~,
\label{rhoxb}
\end{equation}
i.e., the transverse polarization of the proton produces
a shift in the transverse location
of the quarks. 
 As explained before,
this effect in the partonic structure of transversely
polarized protons has been related, in peculiar models,
in a qualitative way, to
a nonvanishing Sivers effect \cite{burk2,die}.

\section{The Sivers function in constituent quark models}

The constituent quark, one of the most fruitful concepts in 20th
century physics, was proposed to explain the structure of the large
number of baryons being discovered in the sixties \cite{quarks}. The
constituent quark concept was incorporated into a QCD scheme by
taking into account gluon exchanges \cite{ruju}. 
The chosen description was a potential model in order
to establish an immediate connection with all previous work.

The constituent quark scheme has guided some of the most successful
parameterizations of parton distributions \cite{gluu}.
Besides, the philosophy that has guided these parameterizations
is precisely the
one used to establish the link between constituent models and parton
distributions. More specifically, model calculations are ascribed to
a scale determined by their partonic content \cite{pape,jaro}. In
most models that scale is characterized by the existence of 
valence quarks only. From that low scale one uses DGLAP evolution to
describe the partonic regime \cite{trvv}.

The models based on constituent quarks (CQMs) have produced
beautiful results in the description of PDs and GPDs, leading to a
phenomenological understanding of them 
in terms of momentum densities and wave
functions \cite{trvv,h1,oam,epj,bpt}. This success in the
description of many parton distributions makes us confident that
the application of the approach to 
the calculation of the Sivers function will also serve to guide the
experimental observations and help the physical interpretation of
this observable.

Let us specify in detail the scheme in which we are going to develop
our formalism for the Sivers function. We shall assume that the
nucleon at a certain low energy scale  is
made up of valence quarks only. These valence quarks are held
together by a confining interaction; in addition, there is a residual
interaction,
governed by the structure of perturbative QCD, e.g. the
One Gluon Exchange Interaction. The 
strong confining interaction maintains the quarks together,
while the residual one
governs the splittings within the same flavor multiplet. Any
scheme with these hypothesis is a constituent quark model framework.

This scheme has never to be understood in a trivial perturbative
sense. The parameters absorb much of the non perturbative features
of the dynamics and this relation between the parameters and
some chosen observables makes the scheme predictive. If one
goes to higher order in the perturbative expansion, one needs to find
new parameters from the chosen observables. Thereafter, the
predictions do not change much with respect to the lowest order
result \cite{maxwell}. Certainly we are dealing with models and not
with QCD and therefore one should not expect precision.
Nevertheless, the scheme has been so successful that particles which
do not fit approximately under it are called exotics, hybrids or
other peculiar names.

Using this scheme we evaluate a formula for the Sivers function,
defined according to Eq. (\ref{work}), valid for any CQM. Let us
proceed to the analysis having in mind Fig. 1. To the first non
vanishing order giving a contribution to the asymmetry, the Sivers
function for the flavor ${\cal Q}$ is obtained as follows:
\begin{eqnarray}
f_{1T}^{\perp {\cal Q}} (x, {k_T} ) =
\Im \left \{
{M \over 2 k_x}
\int
{ d \xi^- d^2 \vec{\xi}_T \over (2 \pi)^3 }
e^{-i ( x \xi^- P^+  - \vec{\xi}_T \cdot {{\vec k_T}} )}
\langle \hat O^{\cal Q} \rangle \right \}~,
\end{eqnarray}
where
\begin{eqnarray}
\langle \hat O^{\cal Q} \rangle & = &
\langle P S_z = 1 |
\bar \psi_{{\cal Q} i}(0,\xi^-,\vec{\xi}_T)
(ig)
\left \{
\hat O_a (0,\xi^-,\vec{\xi}_T)
T^a_{ij}
\right \}
\nonumber \\
& \times &
\gamma^+
\psi_{{\cal Q} j }(0)
| P S_z = - 1 \rangle + \mbox{h.c.}~,
\label{op}
\end{eqnarray}
where $T_{ij}^a = \lambda_{ij}^a/2$ with $\lambda_{ij}^a$
being a Gell-Mann matrix,
and
\begin{eqnarray}
\hat O_a (0,\xi^-,\vec{\xi}_T) & = &
\int_{\xi^-}^\infty A_a^+(0,\eta^-,\vec{\xi}_T) d \eta^-
%e^{i \hat P \cdot \xi}
%\hat O_a(0,0) e^{-i \hat P \cdot \xi}
\nonumber \\
& = &
e^{i \hat P^+ \xi^-
-i \hat P_T \cdot \vec{\xi}_T}
\hat O_a(0)
e^{-i \hat P^+ \xi^-
+i \hat P_T \cdot \vec{\xi}_T}~.
\label{oij}
\end{eqnarray}

In the above equations, use is made of light-cone states\footnote{Here and in the following,
$\tilde x = (x^+,x^-,\vec x_T)$
is a four vector in light-cone coordinates, while obviously
$\vec x = (x_1,x_2,x_3)$ and $\vec x_T = (x_1,x_2)$.},
defined
as $| \tilde p \rangle = | p^+, \vec p_T \rangle$, with
$p^- = (m^2 + p_T^2)/(2 p^+)$.
The light-cone states are
normalized as follows:
\begin{eqnarray}
\langle \tilde p' r' | \tilde p r \rangle =
(2 \pi)^3 2 p^+ \delta (p'^+ - p^+)
\delta(\vec p'_T - \vec p_T)
\delta_{rr'}~,
\label{s-h}
\end{eqnarray}
where the label $r$ represents a set of discrete quantum numbers.
The creation and annihilation
operators of the quark fields are normalized accordingly:
\begin{eqnarray}
\{ b_l^\dagger(\tilde p), b_{l'}(\tilde p') \} =
(2 \pi)^3 2 p^+ \delta (p'^+ - p^+)
\delta(\vec p'_T - \vec p_T)
\delta_{ll'}~,
\end{eqnarray}
where the set $l=\{ m,c,{\cal F} \}$ includes the helicity, color
and flavor quantum numbers of the quark, respectively.

Using the approximation of
expanding Eq. (\ref{op}) in terms of free quark fields \cite{epj},
one gets
\begin{eqnarray}
f_{1T}^{\perp {\cal Q}} (x, k_T ) & = &
\Im \left \{
{
M \over 2 k_x
}
\int
{ d \xi^- d^2 \vec{\xi}_T \over (2 \pi)^3 }
e^{-i ( x \xi^- P^+  - \vec{\xi}_T \cdot {{\vec k_T}} )}
\langle P r S_z = 1 |
\right.
\nonumber \\
& \times &
% \int { d k_1^+ d k_T \over (2 \pi)^3 }
\left.
\int d \tilde k_3
\sum_{m_3}
b_{m_3 i}^{{\cal Q} \dagger} (\tilde k_3) e^{ik_3^+\xi^-
- i \vec k_{3T} \cdot \vec{\xi}_T} \bar u_{m_3}(\vec k_3)
\right.
\nonumber \\
& \times &
\left.
(ig)
%\sum_{l_n,l_1}
%%\int { d \vec{k_n} \over (2 \pi)^3 }
%%\int { d \vec{k_3} \over (2 \pi)^3 }
%\int { d \tilde{k_n}}
%\int { d \tilde{k_1}}
%|\tilde k_1 l_1 \rangle | \tilde k_n l_n \rangle
%\langle \tilde k_n l_n | \langle \tilde k_1 l_1 |
%\right.
%\nonumber \\
%& \times &
%\left.
\left \{
\hat O_a (0,\xi^-,\vec{\xi}_T)
T^a_{ij}
\right \} \gamma^+
\right.
\nonumber \\
%& \times &
%\left.
%\sum_{l_n',l_1'}
%%\int { d \vec{k_n'} \over (2 \pi)^3 }
%%\int { d \vec{k_4} \over (2 \pi)^3 }
%\int { d \tilde{k_n'}}
%\int { d \tilde{k_1'}}
%|\tilde k_1' l_1' \rangle | \tilde k_n' l_n' \rangle
%\langle \tilde k_n' l_n' | \langle \tilde k_1' l_1' |
%\, \gamma^+
%\right.
%\nonumber \\
& \times &
\left.
\sum_{m_3'}
\int { d \tilde{k_3'}}
\, b_{m_3' j }^{\cal Q}
(\tilde k_3') u_{m_3'}(\vec k_3') |P r S_z = -1 \rangle
+ h.c. \right \}~,
\label{muy}
\end{eqnarray}
where $d \tilde k_i = d k^+_i d \vec k_{Ti} / ( 2 k^+_i (2 \pi)^3 )$.
Inserting now proper complete sets of intermediate free one quark
states, the previous equation becomes
\begin{eqnarray}
f_{1T}^{\perp {\cal Q}} (x, k_T ) & = &
\Im \left \{
{
M \over 2 k_x
}
\int
{ d \xi^- d^2 \vec{\xi}_T \over (2 \pi)^3 }
e^{-i ( x \xi^- P^+  - \vec{\xi}_T \cdot {{\vec k_T}} )}
\langle P r S_z = 1 |
\right.
\nonumber \\
& \times &
% \int { d k_1^+ d k_T \over (2 \pi)^3 }
\left.
\int d \tilde k_3
\sum_{m_3}
b_{m_3 i}^{{\cal Q} \dagger} (\tilde k_3) e^{ik_3^+\xi^-
- i \vec k_{3T} \cdot \vec{\xi}_T} \bar u_{m_3}(\vec k_3)
\right.
\nonumber \\
& \times &
\left.
(ig)
\sum_{l_n,l_1}
%\int { d \vec{k_n} \over (2 \pi)^3 }
%\int { d \vec{k_3} \over (2 \pi)^3 }
\int { d \tilde{k_n}}
\int { d \tilde{k_1}}
|\tilde k_1 l_1 \rangle | \tilde k_n l_n \rangle
\langle \tilde k_n l_n | \langle \tilde k_1 l_1 |
\right.
\nonumber \\
& \times &
\left.
\left \{
\hat O_a (0,\xi^-,\vec{\xi}_T)
T^a_{ij}
\right \}
\right.
\nonumber \\
& \times &
\left.
\sum_{l_n',l_1'}
%\int { d \vec{k_n'} \over (2 \pi)^3 }
%\int { d \vec{k_4} \over (2 \pi)^3 }
\int { d \tilde{k_n'}}
\int { d \tilde{k_1'}}
|\tilde k_1' l_1' \rangle | \tilde k_n' l_n' \rangle
\langle \tilde k_n' l_n' | \langle \tilde k_1' l_1' |
\, \gamma^+
\right.
\nonumber \\
& \times &
\left.
\sum_{m_3'}
\int { d \tilde{k_3'}}
\, b_{m_3' j }^{\cal Q}
(\tilde k_3') u_{m_3'}(\vec k_3') |P r S_z = -1 \rangle
+ h.c. \right \}~.
\label{muy_int}
\end{eqnarray}
%where $d \tilde k_i = d k^+_i d \vec k_{Ti} / ( 2 k^+_i (2 \pi)^3 )$.

If there is no further interaction within
the recoiling system, one has:
\begin{eqnarray}
\langle \tilde k_n l_n
| \tilde k_n' l_n' \rangle =
(2 \pi)^3 2 k_n^+ \delta (k_n'^+ - k_n^+)
\delta(\vec k_{n'\,T} - \vec k_{n\,T})
\delta_{l_n,l_n'}~,
\label{stic}
\end{eqnarray}
\begin{eqnarray}
\langle P \, r \, S_z=1 & | & \{
b_{m_3 i}^{{\cal Q} \dagger} (\tilde k_3)
|\tilde k_1 l_1 \rangle | \tilde k_n l_n \rangle \}
\nonumber \\
& = &
(2 \pi)^3 2 k_n^+
\delta( P^+ - k_1^+ - k_3^+ - k_{n}^+ )
\delta( \vec P_T - \vec k_{1\,T} - \vec k_{3\,T} - \vec k_{n\,T} )
\delta_{(S_z,r,l_1,l_3,l_n)}
\nonumber \\
& \times &
\langle P \, r \, S_z=1 | \,
\tilde k_3 \{m_3,i,{\cal Q}\};\, \tilde k_1 \{m_1,c_1,{\cal F}_1 \};
\, \tilde P - \tilde k_3 - \tilde k_1,  l_n \rangle
\nonumber \\
& = &
(2 \pi)^3 2 k_n^+
\delta( P^+ - k_1^+ - k_3^+ - k_{n}^+ )
\delta( \vec P_T - \vec k_{1\,T} - \vec k_{3\,T} - \vec k_{n\,T} )
\delta_{(S_z,r,l_1,l_3,l_n)}
\nonumber \\
& \times &
\Psi^{\dagger}_{r \, S_z=1}
\left ( \tilde k_3 \{m_3,i,{\cal Q}\}; \, \tilde k_1 \{m_1,c_1,{\cal F}_1 \};
\, \tilde P - \tilde k_3 - \tilde k_1,  l_n  \right )~.
\end{eqnarray}
%\begin{eqnarray}
%\{
%\langle \tilde k_n' l_n' |  \langle \tilde k_1' l_1' & | &
%b_{m_3'j}^Q(\tilde k_3')
%\}  |  P \, r \, S_z = -1\rangle
%\nonumber \\
%& = &
%\delta( \tilde P - \tilde k_3' - \tilde k_1' - \tilde k_{n}^{'} )
%\delta_{(S_z,r,l_3',l_1',l_n')}
%\nonumber \\
%& \times &
%\langle
%\tilde P - \tilde k_3' - \tilde k_1' ,l_n' ;
%\, \tilde k_1'  \{m_1',c_1',{\cal F}'_1 \};
%\, \tilde k_3' \{m_3',j,Q\} |P \, r \, S_z = -1  \rangle
%\nonumber \\
%& = &
%\delta( \tilde P - \tilde k_3' - \tilde k_1' - \tilde k_{n}^{'} )
%\delta_{(S_z,r,l_3',l_1',l_n')}
%\nonumber \\
%& \times &
%\Psi_{r \, S_z=-1}
%\left (\tilde k_3' \{m_3',j,Q\}; \, \tilde k_1' \{m_1',c_1',{\cal F}'_1 \};
%\, \tilde P - \tilde k_3' - \tilde k_1',  l_n'  \right )~.
%\end{eqnarray}
In the last equation, the definition of the
intrinsic proton wave function,
$\Psi$,
%and of its conjugated, $\Psi^{\dagger}$,
in momentum space\footnote{In the class of models to be later used,
the separation of the center of mass and intrinsic motion
is always possible.},
has been recovered.
In the same equation, the terms
$\delta_{(S_z, r, ...)}$ are showing that all the
discrete quantum numbers of the quarks have to be properly combined
to recover those of the parent proton.
In order to obtain a workable expression for the Sivers function
given by Eq. (\ref{muy_int}),
other three relations have to be used.
One is written using
Eq. (\ref{oij}) and translational invariance:
\begin{eqnarray}
\langle \tilde k_1 l_1 |
\left \{
\hat O_a (0,\xi^-,\vec{\xi}_T)
\right \}
|\tilde k_1' l_1' \rangle & = &
e^{i k^+_1 \xi^-
-i \vec k_{1 T} \cdot \vec{\xi}_T}
\langle \tilde k_1 l_1 |
\left \{
\hat O_a (0)
\right \}
|\tilde k_1' l_1' \rangle
e^{-i k'^+_1 \xi^-
+i  \vec k_{1 T}' \cdot \vec{\xi}_T}~.
\label{shift}
\end{eqnarray}

Another one is the identity \cite{bjy}:
\begin{eqnarray}
\hat O_a (0)  =
\int_0^\infty A_a^+(\eta^-,0_T) d \eta^- & = &
- \int
{d^4 q \over (2 \pi)^4}
{i \over {q^+ - i \epsilon}}
A_a^+(q)~.
\label{a+}
\end{eqnarray}

The last one is obtained by evaluating the matrix element of the
perturbative free
gluon operator appearing in Eq. (\ref{shift}). Assuming, as an approximation,
that this operator is time-independent, one gets, in the Landau gauge:
\begin{eqnarray}
\langle \tilde k_1 l_1 |
A_a^+(q)
|\tilde k_1' l_1' \rangle & = &
{g \over q^2 } T^a_{c_1c'_1}
\bar u_{m_1}(\vec k_1)
\gamma^+
u_{m_1'}(\vec k_1') \delta_{ {\cal F} {\cal F'}} (2 \pi) \delta(q_0)
\nonumber
\\
& \times &
(2 \pi)^3 2 k_1^+
\delta(k_1^+ - k_1'^+ - q^+ )
\delta(\vec k_{1\,T} - \vec k_{1\,T}' - \vec q )~.
\label{gl}
\end{eqnarray}
Substituting in Eq. (\ref{muy_int}) the identity:
\begin{eqnarray}
{1 \over q^+ - i \epsilon }
-
{1 \over q^+ + i \epsilon } = i ( 2 \pi ) \delta( q^+ )~,
\end{eqnarray}
together with Eqs. (\ref{stic}) -- (\ref{gl}),
one is left with the following expression for the Sivers
function:
\begin{eqnarray}
f_{1T}^{\perp {\cal Q}} (x, {k_T} ) & = &
\Im
\left \{
-i g^2
{
M \over 2 k_x
}
\int
%{ d k_1^+ d k_T \over (2 \pi)^3 }
d \tilde k_1
%\int { d \vec{k_3} \over (2 \pi)^3 }
d \tilde k_3
{d^4 q \over (2 \pi)^3} \delta(q^+)
\right.
\nonumber \\
& \times &
\left.
\delta(k_3^+  + q^+ - xP^+)
\delta(\vec k_{3 T} + \vec q_T - \vec k_T)
{ (2 \pi) \delta(q_0)}
\right.
\nonumber \\
& \times &
\left.
\sum_{{\cal F}_1,m_1,c_1,m_1',c_1',m_3,i,m_3',j}
\delta_{(S_z,r,m_3',m_1',l_n,m_3,m_1,i,j,c_1,c_1')}
\right.
\nonumber \\
& \times &
\left.
\Psi^{\dagger}_{r \, S_z=1}
\left ( \tilde k_3 \{m_3,i,{\cal Q}\}; \, \tilde k_1 \{m_1,c_1,{\cal F}_1 \};
\, \tilde P - \tilde k_3 - \tilde k_1,  l_n  \right )
\right.
\nonumber \\
& \times &
\left.
T^a_{ij}  T^a_{c_1c_1'}
V(\vec k_1, \vec k_3, \vec q)
\right.
\nonumber \\
& \times &
\left.
\Psi_{r \, S_z=-1}
\left (\tilde k_3+ \tilde q, \{m_3',j,{\cal Q}\}; \,
\tilde k_1 - \tilde q, \{m_1',c_1',{\cal F}_1 \};
\, \tilde P - \tilde k_3 - \tilde k_1,  l_n  \right )
\right \}~,
\label{start}
\end{eqnarray}

with the interaction determined by:

\begin{eqnarray}
V(\vec k_1, \vec k_3, \vec q)=
{ 1 \over q^2}
\bar u_{m_3}(\vec k_3) {\gamma^+ }
u_{m_3'}(\vec k_3 + \vec q)
\bar u_{m_1}(\vec k_1) \gamma^+
u_{m_1'}(\vec k_1 - \vec q)~.
\label{pot}
\end{eqnarray}

Since the final aim is the evaluation of the Sivers function within
a NR model, a NR reduction of the interaction
has to be performed. 
Using therefore
the definitions of free four-spinors in Eq. (\ref{pot}),
performing a NR expansion leaving out terms of second order in momentum,
as it is commonly done in nuclear physics (cf. Ref. \cite{mach}), one gets
the potential:
\begin{eqnarray}
V_{NR}(\vec k_1, \vec k_3, \vec q) & =
& \frac{1}{2q^2} \left [ (V_0)_{m_1,m_1',m_3,m_3'} + (V_S)_{m_1,m_1',m_3,m_3'} \right]~,
\label{potential}
\end{eqnarray}
with:
\begin{eqnarray}
V_0(\vec k_1, \vec k_3, \vec q)_{m_1,m_1',m_3,m_3'} =\left[
1 + {k_3^z \over m} + { \vec q \cdot \vec k_3 \over 4 m^2 }
+ {k_1^z \over m} - { \vec q \cdot \vec k_1 \over 4 m^2 }
+ O \left ( { k_1^2 \over m^2},{ k_3^2 \over m^2} \right )\right]
\delta_{m_1,m_1'}\delta_{m_3,m_3'}
\label{pot0}
\end{eqnarray}
\begin{eqnarray}
V_S(\vec k_1, \vec k_3, \vec q)_{m_1,m_1',m_3,m_3'} & = &
-i 
\left (
1 + {k_3^z \over m} + { \vec q \cdot \vec k_3 \over 4 m^2 } \right )\delta_{m_3,m_3'}
{ [\vec q \times (\vec \sigma_1)_{m_1,m_1'}]_z \over 2  m}
\nonumber
\\
&&+ i 
\left (
1 + {k_1^z \over m} - { \vec q \cdot \vec k_1 \over 4 m^2 } \right )\delta_{m_1,m_1'}{ [ \vec q  \times(\vec\sigma_3)_{m_3,m_3'} ]_z \over 2 m }
\nonumber
\\
&& +  i \delta_{m_1,m_1'}{  (\vec\sigma_3)_{m_3,m_3'}  \cdot (\vec k_3 \times \vec q) \over 4 m^2 }\nonumber
\\
&&- i {  (\vec\sigma_1)_{m_1,m_1'} \cdot (\vec k_1 \times \vec q) \over 4 m^2 }\delta_{m_3,m_3'}
\nonumber
\\
& &+  { 
[\vec q \times (\vec\sigma_1)_{m_1,m_1'} ]_z (\vec\sigma_3)_{m_3,m_3'}  
\cdot (\vec k_3 \times \vec q)
\over 8 m^3 }\nonumber
\\
&&+ {  (\vec\sigma_1)_{m_1,m_1'}  \cdot (\vec k_1 \times \vec q)
[\vec q \times (\vec\sigma_3)_{m_3,m_3'} ]_z\over 8 m^3
 }
\nonumber
\\
& &+  {[\vec q \times (\vec\sigma_1)_{m_1,m_1'} ]_z 
[\vec q \times(\vec\sigma_3)_{m_3,m_3'} ]_z 
\over 4 m^2 } + O \left ( { k_1^2 \over m^2},{ k_3^2 \over m^2} \right )~.
\label{vnr}
\end{eqnarray}
A few remarks are in order. First of all, the helicity conserving
part, $V_0$, Eq. ({\ref{pot0}}), of the global interaction
Eq. ({\ref{potential}}), does not contribute to the Sivers function.
One should notice that, in an extreme NR limit, the Sivers function
would turn out to be identically zero.
In our approach, it is precisely the interference of the small
and large components in the four-spinors
of the free quark states which leads to a non-vanishing
Sivers function, even from
the component with $l = 0$ of the target wave function.
Effectively, these interference terms in the interaction are
the ones that, in other approaches, arise due to the wave function
(see, e.g., the MIT bag model calculation \cite{yuan}).
% Of course, all the components with $l \neq 0$ yield a contribution.

The scheme is now completely set up and any CQM can be used to 
evaluate the Sivers function. 
We next use  properly normalized NR wave functions to transform
%Besides, since the calculation will be performed in a NR
%model,
%properly normalized states $ | \vec p r \rangle$
% and creation and annihilation operators $b(\vec p)$
%have to be used, i.e.
%\begin{eqnarray}
%\langle \vec p' \, r' | \vec p \, r \rangle =
%\delta(\vec p' - \vec p) \delta_{rr'}~.
%\label{nornr}
%\end{eqnarray}
%%\begin{eqnarray}
%%\{
%%b^\dagger( \vec p),
%%b( \vec p') \} = \delta( \vec p - \vec p')~.
%%\label{bnr}
%%\end{eqnarray}
%In terms of these,
 Eq. (\ref{start}) in:
\begin{eqnarray}
f_{1T}^{\perp {\cal Q}} (x, {k_T} ) & = &
\Im
\left \{
- i g^2
{
M^2 \over k_x
}
\int
%{ d k_1^+ d k_T \over (2 \pi)^3 }
d \vec k_1
%\int { d \vec{k_3} \over (2 \pi)^3 }
d \vec k_3
{d^2 \vec q_T \over (2 \pi)^2}
\delta(k_3^+ - xP^+)
\delta(\vec k_{3 T} + \vec q_T - \vec k_T) {\cal M}^{\cal Q}
\right \}~,
\label{start2}
\end{eqnarray}
where
\begin{eqnarray}
{\cal M}^{\cal Q} & = &
\sum_{{\cal F}_1,m_1,c_1,m_1',c'_1,m_3,i,m_3',j}
\delta_{(S_z,r,m_3',m_1',l_n,m_3,m_1,i,j,c_1,c'_1)}
\nonumber \\
& \times &
\Psi^{\dagger}_{r \, S_z=1}
\left ( \vec k_3 \{m_3,i,{\cal Q}\}; \, \vec k_1 \{m_1,c_1,{\cal F}_1 \};
\, \vec P - \vec k_3 - \vec k_1,  l_n  \right )
\nonumber \\
& \times &
T^a_{ij}  T^a_{c_1c_1'}
V(\vec k_1, \vec k_3, \vec q)
\nonumber \\
& \times &
\Psi_{r \, S_z=-1}
\left (\vec k_3 + \vec q, \{m_3',j,{\cal Q}\}; \, \vec k_1 -
\vec q, \{m_1',c_1',{\cal F}_1 \};
\, \vec P - \vec k_3 - \vec k_1,  l_n  \right )~.
\label{M}
\end{eqnarray}
Each wave function $\Psi_{r \, S_z}$
describing a possible
proton state can be factorized into a completely
antisymmetric color wave function,
$ \chi$, and a symmetric spin-flavor-momentum
state, $\Phi_{sf}$, as follows:
\begin{eqnarray}
\Psi_{r \, S_z}=
\Phi_{sf,S_z}
\left ( \vec k_3 \{m_3, {\cal Q}\}; \, \vec k_1 \{m_1,{\cal F}_1 \};
\, \vec P - \vec k_3 - \vec k_1,  \{m_n, {\cal F}_n \}  \right )
\chi(i,c_1,c_n)~.
\label{fact}
\end{eqnarray}
The matrix element of the color operator in Eq. (\ref{M})
can be therefore immediately evaluated:
\begin{eqnarray}
\sum_{c_1,c'_1,i,j}
\chi^\dagger(i,c_1,c_n)
T^a_{ij}  T^a_{c_1c_1'}
\chi(j,c_1',c_n)= - { 2 \over 3 }~,
\label{chicol}
\end{eqnarray}
which is the well-known result
for the exchange of one gluon between quarks in a color singlet
3-quark state \cite{mmg}. Besides,
as a consequence of the symmetry of the state $ \Phi_{sf}$,
one can assume that the interacting quark is the one labeled
``3'', so that,
after the evaluation
of the summation on the flavors ${\cal F}_1$,
${\cal M}^\alpha$ can be written, for the $u$
and $d$ flavors, as follows:
\begin{eqnarray}
{\cal M}^{u(d)} & = &
\left ( - { 2 \over 3} \right ) \cdot 3 \cdot
\sum_{m_1,m_1',m_3,m_3'}
\Phi_{sf,S_z=1}^{\dagger}
\left ( \vec k_3, m_3; \vec k_1, m_1;
\, \vec P - \vec k_3 - \vec k_1,  m_n  \right )
%\left (
%{ 1 + \tau_3(1) \over 2 } + { 1 \mp \tau_3(1) \over 2 }
%\right )
\nonumber
\\
& \times &
{ 1 \pm \tau_3(3) \over 2 }
V_{NR}(\vec k_1, \vec k_3, \vec q)
\nonumber
\\
& \times &
\Phi_{sf , S_z=-1}
\left (\vec k_3 + \vec q, m_3'; \, \vec k_1 -
\vec q, m_1';
\, \vec P - \vec k_3 - \vec k_1,  m_n  \right )~.
\label{Mu}
\end{eqnarray}
%and it has been considered that the flavor operator
%$(1 + \tau_3(1) / 2 ) + ( 1 \mp \tau_3(1) / 2 )$,
%where $\pm$ refers to the $u(d)$ flavor,
%yields 1 when evaluated for the proton wave function.
Eq. (\ref{start2}), with ${\cal{M}}^{u(d)}$
given by Eq. (\ref{Mu}),
provides us with a suitable formula
to evaluate the Sivers function,
once the spin-flavor wave function of the proton
in momentum space, i.e. the quantity
$\Phi_{sf}$, is available
in a given constituent quark model.

\section{The calculation of the Sivers function in the Isgur-Karl model}

As an illustration,
in this section we present the results of our approach
in the CQM of Isgur and Karl (IK) \cite{ik}.
In this model
the proton wave function is obtained in a
OGE potential
added to a confining harmonic oscillator (H.O.);
including contributions up to the $2 \hbar \omega$ shell,
 the proton state is given by the
following admixture of states
\begin{eqnarray}
|N \rangle =
a | ^2 S_{1/2} \rangle_S +
b | ^2 S'_{1/2} \rangle_{S} +
c | ^2 S_{1/2} \rangle_M +
d | ^4 D_{1/2} \rangle_M~,
\label{ikwf}
\end{eqnarray}
where the spectroscopic notation $|^{2S+1}X_J \rangle_t$,
with $t=A,M,S$ being the symmetry type, has been used.
The coefficients were determined by spectroscopic properties to be
$a = 0.931$,
$b = -0.274$,
$c = -0.233$, $d = -0.067$
\cite{mmg}.
If $a = 1$ and
$b =
c = d = 0$, a simple H.O.
model
is recovered.
The parameter $\alpha^2 = m \omega$ of the H.O
potential is fixed to the value 1.23 fm$^{-2}$, in order
to reproduce the slope of the proton charge form factor at zero 
momentum transfer
\cite{mmg}.

The formal expressions of the wave functions appearing in Eq. (\ref{ikwf})
in the IK model can be found in \cite{mmg,ik2},
given in terms
of the following sets of
conjugated intrinsic coordinates
\begin{eqnarray}
\vec R = { 1 \over \sqrt{3} } ( \vec{r_1} + \vec{r_2} + \vec{r_3} )
& \leftrightarrow &
\vec K = { 1 \over \sqrt 3} ( \vec{k_1} + \vec{k_2} + \vec{k_3} )
\nonumber ~,
\\
\vec \rho = { 1 \over \sqrt 2} ( \vec{r_1}  - \vec{r_2} )
& \leftrightarrow &
\vec{k_{\rho}} = { 1 \over \sqrt 2} ( \vec{k_1} -  \vec{k_2} )
\nonumber ~,
\\
\vec \lambda = { 1 \over \sqrt 6} ( \vec{r_1} + \vec{r_2} - 2 \vec{r_3} )
& \leftrightarrow &
\vec{k_{\lambda}} = { 1 \over \sqrt 6} ( \vec{k_1} + \vec{k_2} - 2 \vec{k_3} )
~.
\label{coor}
\end{eqnarray}

There are many good reasons to use the IK model as a test
of the developed formalism.
First of all, the IK is the typical CQM,
succesful in reproducing the low-energy properties of the nucleon,
such as the spectrum and the elastic and transition form factors
at small momentum transfer \cite{ik,mmg}.
In particular,
as was shown in Ref.~\cite{Conci:1990kt}, in the IK model, 
$\langle k^2\rangle /m^2 \sim 0.3$ and therefore 
one expects small corrections from terms $O\left ( { k^2 / m^2} \right )$.
Besides, one of the features of the IK model is that the OGE
 mechanism \cite{ruju}, which reduces the degeneracy of the spectrum,
is taken into account. It is therefore natural
to study our formalism, based on OGE FSI,
within the IK framework.
Concerning PDs,
it has been shown that the IK model can describe their gross
features, once QCD evolution of the proper matrix elements
of the corresponding twist-2 operators is performed from the scale
of the model to the experimental one \cite{trvv,h1,oam}.
Reasonable predictions of GPDs have also been obtained \cite{epj},
and this makes particularly interesting the evaluation of the Sivers
function in the IK model.
In section II, the relation between the Sivers function
and the impact parameter dependent GPDs has been discussed.
In a model where a shift
of the quark location in the transverse plane
is found,
a sizable Sivers function should  arise.
In order to investigate
whether the IK model is  suitable  for the
analysis of the Sivers function,
the quantity $\rho_{\cal Q}(x,\xi=0,\vec b)$ has been calculated in this
 model \cite{fratinif}, performing the Fourier transforms,
Eq. (\ref{ipd}), of GPDs evaluated along the lines of Ref.
\cite{epj}.
The quantity:
\begin{equation}
\rho_{\cal Q}(\vec b) = \int dx \tilde \rho_{\cal Q}(x,\xi=0,\vec b)~,
\label{rhob}
\end{equation}
representing
the distribution of the quarks of flavor ${\cal Q}$, with
any longitudinal momentum, in the transverse plane,
independently of their helicity, in a proton polarized along
the positive $y$ direction, is shown in Fig. 2.
It is clear that a slight shift along the $x$ direction is observed,
with a different sign for the $u$ and $d$ flavor.
%a little stronger for the $d$ rather than for the $u$ flavor.
Therefore, according to the present wisdom,
a small Sivers function is expected, with different sign
for the $u$ and $d$ flavors \cite{burk2}.

After this discussion, the IK model appears as a promising framework
for the evaluation of the Sivers function.

The Sivers function has been
calculated according to Eq. (\ref{start2}),
using the proton states Eq. (\ref{ikwf}),
neglecting
the small ${D}$ component,
and the potential Eq. (\ref{potential})
in ${\cal{M}}^{u(d)}$ given by Eq. (\ref{Mu}).

The results of the calculation can be cast in the following form:
\begin{eqnarray}
f_{1T}^{\perp {\cal Q}=u,d} (x, {k_T} )
=
& - &
%{\sqrt{2} g^2 M^2 \over 72 \pi^2 k_x}
{\sqrt{2} g^2 M^2 \over k_x}\left(\frac{3}{2}\right)^{\frac{3}{2} }{1\over 2 \pi^{3/2} \alpha^3}
\int {\frac{d^2 \vec q_T}{(2\pi)^2}} { k_\lambda^0 \over  |k_\lambda^0 -
 k_\lambda^z|}
e^{ - \frac{1}{ \alpha^2 }\left[k_\lambda^2 +\frac{7}{8} q_T^2-\sqrt{\frac{3}{2}} \vec{q}.\vec{k}_{\lambda}\right]}
\nonumber
\\
& \times &
\frac{1}{2m} \Bigg [a^2 \frac{q_x}{q^2} \,p_{SS}^{({\cal Q})}
                  +ab   \frac{q_x}{q^2} \left(p_{S'S}^{({\cal Q})}+p_{SS'}^{({\cal Q})}\right)
                 +ac \frac{q_x}{q^2} \left(p_{MS}^{({\cal Q})}+p_{SM}^{({\cal Q})}\right)\nonumber\\
            &&   +ac\,\left(p_{SM'}^{({\cal Q})}+p_{M'S}^{({\cal Q})}\right)
               +b^2 \,\frac{q_x}{q^2}p_{S'S'}^{({\cal Q})}
   +bc \,\frac{q_x}{q^2} p_{S'M}^{({\cal Q})}+bc \frac{q_x}{q^2} p_{MS'}^{({\cal Q})}\nonumber\\
 &&
   +bc\, \left(p_{S'M'}^{({\cal Q})}+p_{M'S'}^{({\cal Q})}\right)
  +c^2\frac{q_x}{q^2} \left(p_{MM}^{({\cal Q})}+p_{M'M'}^{({\cal Q})}\right)
    +c^2\,\left(p_{MM'}^{({\cal Q})}+p_{M'M}^{({\cal Q})}\right)\Bigg ]~,\nonumber\\&&
\label{result}
\end{eqnarray}
with
$k_\lambda^0 = \sqrt{ m^2 +  k_\lambda^2}$, and
\beq
\vec{k}_{\lambda}=\sqrt{\frac{3}{2}}(\vec{q}-\vec{k})~,&&\quad
k_{\lambda}^z=\frac{\frac{3}{2}m^2 + \vec{k}_{\lambda T}^2
- 3 x^2 P^{+ 2} }{2 \sqrt{3}P^+ x}\nonumber\\
k_{\lambda}^2&=&k_{\lambda}^{z 2}+\vec{k}_{\lambda T}^2\quad.
\eeq 
%$\bar k_\lambda = \sqrt{ \bar k_{\lambda z}^2 + \bar k_{\lambda T}^2}$,
%$\bar k_{\lambda T}^2 = {3 \over 2} (\vec q_T - \vec k_T)^2$.
%and
%\begin{equation}
%\%bar k_{\lambda_z} = {\bar k_{\lambda T}^2 + m^2 - 3 x^2 M^2/2
%\over
%\sqrt{6} x M }~.
%\end{equation}
The expressions of the functions $p_{XX}^{({\cal Q})}$ 
are given in the Appendix.

%It is important to notice that the term contributing
%to the Sivers function, in the potential Eq. (\ref{vnr}),
%is the spin-orbit-like coupling
%\begin{equation}
%V_{SO}^{i=1,3} =  { ( \vec q  \wedge \vec \sigma_{i})_z \over 2 m }
%\label{s-o}
%\end{equation}
%between the orbital angular momentum of the exchanged gluon
%and the spin of the quark.
%In particular, only processes with a single spin-flip
%of one of the two interacting quarks give a contribution.
%The same is found in other model estimates, for example
%in Ref. \cite{yuan}.

To evaluate numerically Eq. (\ref{result}), the
strong coupling constant $g$, and therefore
$\alpha_s(Q^2)$, has to be fixed.
%There are two strong coupling constants in our calculation. One, $\alpha$, is associated with the wave function and corresponds to the strength 
%of the interaction at the hadronic scale. The other, $\alpha_s$, is associated with the interaction and runs with the evolution equations. The latter is determined
%by the parton content at the hadronic scale. In a scheme to all orders, both couplings should be equal at the hadronic scale. At present they are close in value but not equal.
Here, the prescription introduced in the past for calculations of
PDFs in quark models (see, i.e., Ref. \cite{trvv}) will be used.
It consists in fixing the momentum scale of the model,
the so-called hadronic scale $\mu_0^2$, according to the
amount of momentum carried by the valence quarks in the model.
In the approach under scrutiny, only valence quarks contribute.
Assuming that all the gluons and sea pairs in the proton
are produced perturbatively according to NLO evolution equations,
in order to have $\simeq 55 \% $ of the momentum
carried by the valence quarks at a scale of 0.34 GeV$^2$,
as in
typical low-energy parameterizations \cite{gluu}, one finds,
that $\mu_0^2 \simeq 0.1$ GeV$^2$
if $\Lambda_{QCD}^{NLO} \simeq 0.24$ GeV.
This yields $\alpha_s(\mu_0^2)/(4 \pi) \simeq 0.13$ \cite{trvv}.

For an easy presentation,
the quantity which is usually shown for the results of calculations
or for data of the Sivers function is its first moment, defined
as follows :
\begin{equation}
f_{1T}^{\perp (1) {\cal Q} } (x)
= \int {d^2 \vec k_T}  { k_T^2 \over 2 M^2}
f_{1T}^{\perp {\cal Q}} (x, {k_T} )~.
\label{momf}
\end{equation}

The results of the present approach
for the moments Eq. (\ref{momf}) are given
by the dashed curves in Fig. 3 (4) for
the $u$ ($d$) flavor.
They are compared
with a parameterization of the HERMES data,
corresponding to an experimental scale of $Q^2=2.5$ GeV$^2$
\cite{coll3}\footnote{It has been chosen
to compare the results with the parameterization
of \cite{coll3} and not with that of \cite{ans} or
\cite{Vogelsang:2005cs}
just because, in the first case, it is easier to reconstruct the
parameterization of the data, 
and their 1-sigma range 
has been kindly provided by the authors
of Ref. [24]. 
The discussion of the quality of the agreement of the present
results with data would not change substantially if the comparison
were made with the parameterization of Refs. \cite{ans,Vogelsang:2005cs}.}
The patterned area represents the $1-\sigma$ range
of the best fit proposed in Ref. \cite{coll3}.

As expected from the IPD GPDs analysis, shown in Fig.~\ref{filippo},
a different sign for the $u$ and $d$ flavor is found.

Let us see now how the results of the calculation
compare with the Burkardt sum rule
\cite{Burkardt:2004ur}, 
which follows from general principles
and must be satisfied at any scale.
If the proton is polarized in the positive $y$ direction,
in our case, where only valence quarks are present,
the Burkardt sum rule reads:
\begin{equation}
\sum_{{\cal Q}=u,d} \langle k_x^{\cal{Q}} \rangle = 0~,
\label{burksr}
\end{equation}
where
\begin{equation}
\langle k_x^{\cal{Q}} \rangle = - \int_0^1 d x \int d \vec k_T
{k_x^2 \over M}  f_{1T}^{\perp \cal{Q}} (x, {k_T} )~.
\label{burs}
\end{equation}
Within our scheme, at the scale of the model, it is found
$\langle k_x^{u} \rangle = 10.85\, MeV$,
$\langle k_x^{d} \rangle = - 11.25\, MeV$ and, 
in order to have an estimate
of the quality of the agreement of our results with
the sum rule, we define the ratio
\begin{equation}
r= {
\langle k_x^{d} \rangle+
\langle k_x^{u}\rangle
\over
\langle k_x^{d} \rangle-
\langle k_x^{u} \rangle}~,
\label{rbur}
\end{equation}
obtaining $r \simeq 0.02$, so that we can say that our calculation
fulfills the Burkardt sum rule to a precision of a few percent.

Another prediction has been derived in the framework of large $N_c$
\cite{pob}
and it reads,
when $x N_c\sim O(1)$  and the large $N_c$ predictions
are supposed to be applicable:
\begin{equation}
r_{NC}=
{
| f_{1T}^{\perp (1) u } (x) + f_{1T}^{\perp (1) d } (x)  |
\over
| f_{1T}^{\perp (1) u } (x) - f_{1T}^{\perp (1) d } (x)  |
}
\simeq { 1 \over N_c }~.
\label{rnc}
\end{equation}
We get the closest value to the
prediction above, 0.26, in a narrow region around $x=0.4$. 

We note that the contribution of the states $|^2 S'_{1/2}\rangle_S$ and $|^2 S_{1/2}\rangle_M$, in spite of their small probability in the proton state
 Eq.~(\ref{ikwf}), turns out to be important in the evaluation of the Sivers function. 

The magnitude of the results
is close to that of the data,
although
they have a different shape: the maximum (minimum)
is predicted at larger values of $x$.
One should anyway realize that one step of the analysis
is still missing: the scale of the model,
$\mu_0^2$, is much lower than the one of the
data, which is $Q^2 =2.5$ GeV$^2$.
For a proper comparison, the QCD
evolution from the model scale to the experimental one
would be necessary. This issue is discussed
in the next session.

\section{QCD evolution of the model
calculation}

The Sivers function is a TMD PDs and the evolution
of this class of functions is, to a large
extent, still
to be understood.
In any case, recent interesting developements can
be found in Ref. \cite{cecco}.

In order to have an indication of the effect of the
evolution, we perform a NLO evolution of the model
results assuming, for the moments of the Sivers function,
the ones defined in Eq. (\ref{momf}),
the same anomalous dimensions of the unpolarized PDFs.
%For the Non Singlet (NS) sector, anomalous dimensions have been
%calculated in the large $N_c$ approximation, at LO
%\cite{henn}.
%The evolution of the results previously described for
%the Sivers function has been
%performed using this approximation.
As described in the previous section,
the parameters of the evolution have been fixed
in order to have a fraction $\simeq 0.55$ of the momentum
carried by the valence quarks at 0.34 GeV$^2$, as in
typical parameterizations of PDFs \cite{gluu}, starting from
a scale of $\mu_0^2 \simeq 0.1$ GeV$^2$ with only valence quarks.
The final result is given by the full curve in Fig. 3 (4)
for the $u$ ($d$) flavor.
As it is clearly seen, the agreement with data
improves dramatically and their
trend is reasonably reproduced at least for $x \ge 0.2$.

Of course a word of caution is in order:
the performed evolution is not really correct.
In any case, an indication of two very important things
is obtained:

i) The evolution of the model result is necessary to estimate
the quantities at the momentum scale of experiments,
as it happens for standard PDs \cite{trvv,h1,oam};

ii) after evolution, the present calculation could be
consistent with data,
at least with the present ones, still
affected by large statistical and systematic errors.

\section{Conclusions}

A rather general formalism for the evaluation
of the Sivers function, to be used in any CQM,
has been developed.
The crucial ingredient has been the
NR reduction of the leading twist part
of the OGE diagram in the final state.
It has been shown that the IK model, based
also on a OGE contribution to the Hamiltonian,
is a proper framework for the estimate of the Sivers function.
The obtained results show a sizable effect, with an opposite
sign for the $u$ and $d$ flavors.
This is in agreement with
the pattern found from an analysis of impact parameter dependent
GPDs in the IK model.

Let us compare our approach with previous calculations.
The diquark model with scalar diquarks has no contribution for the $d$-quark \cite{bacch}  and therefore does not satisfy the Burkardt sum rule (BSR), Eq.~(\ref{burksr}).
The diquark model with axial-vector diquarks has contributions to both $u$ and $d$-quarks and with opposite sign, but with the magnitude of the $d$ 10 times smaller than that of the $u$. The BSR is not satisfied.
The MIT bag model calculation \cite{yuan} has non-vanishing $u$ and $d$-quarks contribution of opposite sign which are proportional in magnitude. The $d$-quark contribution is much smaller than ours and therefore does not satisfy the BSR.
The MIT bag model modified by instanton effects \cite{d'a} has $u$ and $d$-quark contributions of the same sign and therefore does not satisfy the BSR. As a summary, we can say that our calculation, despite the naive wave function used, is in better agreement with the data
with respect to the other approaches, and fulfills the BSR.

In order to compare with the data, one has to evolve the model calculation to the experimental scale.
Although a consistent QCD evolution of the model results to the experimental
momentum scale is not yet possible, due to the lack of the
calculation of the corresponding anomalous dimensions,
an estimate of the evolution has been attempted.
It has been found that, once properly evolved, the model results
could be in reasonable agreement with the available data.

The formalism presented here can be used with any CQM
and it will be interesting in the near future to
implement other calculations with different models, performing
a correct evolution as soon as the corresponding ingredients
become available. The connection of the Sivers function with
IPD GPDs deserves a careful analysis and will be discussed elsewhere.

\vskip 1.cm

\appendix*

\section{The Sivers function in the IK model}

The functions $p_{XX}^{({\cal Q})}$ appearing in Eq.~(\ref{result}) are listed below. 
%\beq
%\vec{k}_{\lambda}=\sqrt{\frac{3}{2}}(\vec{q}-\vec{k})~,&&\quad
%k_{\lambda}^z=\frac{\frac{3}{2}m^2 +\frac{3}{2} \vec{k}_{\lambda}^2- 3 x^2 P^{+ 2} }{2 \sqrt{3}P^+ x}\nonumber\\
%k_{\lambda}^2&=&k_{\lambda}^{z 2}+\vec{k}_{\lambda}^2\quad.
%\eeq 
\beq
p_{SS}^{(u)} &=&(A-\frac{q^2}{18 m^2})\quad,\nonumber\\
p_{SS}^{(d)}  &=&(B+\frac{q^2}{72 m^2})\quad,\\
 p_{S'S}^{(u)}&=&\frac{1}{\sqrt{3}\alpha^2 }
 	 \left[A \left(\frac{3}{2}\alpha^2+\frac{q^2}{8}\right)-
     	  5 \alpha^2\frac{q^2}{36 m^2}-\frac{q^4} {144 m^2}+(A-\frac{q^2}{18 m^2})
           (k_{\lambda}^2-3 \alpha^2)\right]\quad,\nonumber\\
p_{S'S}^{(d)}&=&\frac{1}{\sqrt{3}\alpha^2 }
       \left[B\left(\frac{3}{2}\alpha^2+\frac{q^2}{8}\right)+
      5\alpha^2 \frac{q^2}{144 m^2}+\frac{q^4}{576 m^2}+ (B+\frac{q^2}{72 m^2})
  (k_{\lambda}^2-3 \alpha^2)\right]\quad,\\
 p_{SS'}^{(u)}&=&\frac{1}{\sqrt{3} \alpha^2 }\Big[A \left(\frac{3}{2}\alpha^2+\frac{q^2}{8}\right)-
     	  5 \alpha^2\frac{q^2}{36 m^2}-\frac{q^4} {144 m^2}+(A-\frac{q^2}{18 m^2})\nonumber\\
   &&
           (k_{\lambda}^2-3 \alpha^2+2 q^2 -3 \vec{q}\cdot(\vec{q}-\vec{k}))-A \frac{q^2}{2}+\frac{q^4}{36 m^2}+\alpha^2 \frac{q^2}{9 m^2}\Big]\quad,\nonumber\\
   p_{SS'}^{(d)}&=&\frac{1}{\sqrt{3}\alpha^2 } \Bigg[B\left(\frac{3}{2}\alpha^2+\frac{q^2}{8}\right)+
     5\alpha^2 \frac{q^2}{144 m^2}+\frac{q^4}{576 m^2}+ (B+\frac{q^2}{72 m^2})\nonumber\\
      &&
  (k_{\lambda}^2-3 \alpha^2+2q^2-3 \vec{q}\cdot(\vec{q}-\vec{k})) -B \frac{q^2}{2}-\alpha^2 \frac{q^2}{36 m^2}-\frac{q^4}{144 m^2}\Bigg]\quad,\\
  p_{MS}^{(u)}&=&\frac{1}{\sqrt{6} \alpha^2} \Bigg[-k_{\lambda}^2 \left(D-\frac{5q^2}{72 m^2}\right)+ D \left(\frac{3}{2}\alpha^2+\frac{q^2}{8}\right)
      -25\alpha^2 \frac{q^2}{144 m^2}-5 \frac{q^4}{576 m^2}\Bigg]\quad,\nonumber\\
  p_{MS}^{(d)}&=&\frac{2}{\sqrt{6}\alpha^2 } \Big[k_{\lambda}^2 \left(B+\frac{q^2}{72 m^2}\right)-
    B \left(\frac{3}{2}\alpha^2+\frac{q^2}{8}\right)-5\alpha^2 \frac{q^2}{144 m^2}- \frac{q^4}{576 m^2}\Big]\quad,\\     
 p_{SM}^{(u)}&=&p_{MS}^{(u)}+\frac{1}{\sqrt{6}\alpha^2}\Bigg[ -(q^2-\sqrt{6} \vec{q}\cdot\vec{k}_{\lambda})
\left(D-\frac{5 q^2}{72 m}\right)+\left(-D\frac{q^2}{2}+5 \alpha^2 \frac{q^2}{36 m^2}+ 5 \frac{q^4}{144 m^2}\right)\Bigg]\quad,\nonumber\\
  p_{SM}^{(d)}&=&p_{MS}^{(d)}+\frac{1}{\sqrt{6} \alpha^2 } \Bigg[\left(q^2-\sqrt{6} \vec{q}\cdot\vec{k}_{\lambda}\right) \left(B+\frac{q^2}{72 m^2}
   \right) 
   +\left(B \frac{q^2}{2}+\frac{\alpha^2 q^2}{36 m}+ \frac{q^4}{144 m^2}\right)\Bigg]\quad,\\
  p_{M'S}^{(u)}&=&-\frac{2}{ q^2\sqrt{18}\alpha^2}
           \Bigg[ -\alpha^2 q_x  \frac{k_{\lambda}^z}{4 \sqrt{2} m}
              -q_y \alpha^2 \frac{(\vec{k} \times \vec{q})^z}{8 \sqrt{2} m^2}
     +\frac{\vec{q}\cdot\vec{k}_{\lambda}}{2 \sqrt{2}} \,C_{MS MA}^{M'S}+\nonumber\\
      && \frac{1}{3}\left(-\alpha^2 q_x \frac{k_{\lambda}^z}{4 \sqrt{2} m}+q_y \alpha^2\frac{(\vec{k} \times \vec{q})^z}{8 \sqrt{2} m^2} +
\frac{\vec{q}\cdot\vec{k}_{\lambda}}{2 \sqrt{2}} \,C_{MA MS}^{M'S}\right)\Bigg]\quad,\nonumber\\
         p_{M'S}^{(d)}&=&-\frac{2}{ q^2\sqrt{18}\alpha^2}\frac{2}{3} \Bigg[-\alpha^2 q_x \frac{k_{\lambda}^z}{4 \sqrt{2} m}+q_y
   \alpha^2 \frac{(\vec{k} \times \vec{q})^z}{8 \sqrt{2} m^2}+  \frac{\vec{q}\cdot\vec{k}_{\lambda}}{2 \sqrt{2}}\, C_{MA MS}^{M'S}\Bigg]\quad,\\
        p_{SM'}^{(u)}&=&  p_{M'S}^{(u)} -\frac{2}{q^2 \sqrt{18}\alpha^2 }\left(C_{MS MA}^{M'S}+\frac{1}{3}C_{MA MS}^{M'S}\right)\left( \frac{\sqrt3 q^2}{4}-\frac{\vec{q}\cdot\vec{k}_{\lambda}}{\sqrt{2}}\right)\quad,\nonumber\\
        p_{SM'}^{(d)}&=&  p_{M'S}^{(d)}   -\frac{2}{q^2 \sqrt{18}\alpha^2 }\frac{2}{3} C_{MA MS}^{M'S} \,\left( \frac{\sqrt3 q^2}{4}-\frac{\vec{q}\cdot\vec{k}_{\lambda}}{\sqrt{2}} \right)\quad,\\
  p_{S'S'}^{(u)} &=&\frac{1}{3 \alpha^4}\Bigg[  F \, p_{SS}^{(u)}  +G \left(A \left(\frac{3}{2}\alpha^2+\frac{q^2}{8}\right)
   - 5 \alpha^2 \frac{q^2}{36 m^2} -\frac{q^4}{ 144 m^2}\right)+H \Big( -A\, \frac{q^2}{2}\nonumber\\
     &&+\frac{q^4}{36 m^2}+\alpha^2 \frac{q^2}{9 m} \Big)
       -\sqrt{2} \left(C_{MA}^{S'S' (1)}+\frac{1}{3} C_{MS}^{S'S' (1)}\right)   +\left(C_{MA}^{S'S' (2)}+\frac{1}{3}C_{MS}^{S'S' (2)}\right)\Bigg]\quad,\nonumber\\
       p_{S'S'}^{(d)} &=&\frac{1}{3 \alpha^4 }\Bigg[  F\,  p_{SS}^{(d)} +G \left(B \left(\frac{3}{2}\alpha^2+\frac{q^2}{8}\right)+
     5 \alpha^2 \frac{q^2}{144 m^2} +\frac{q^4}{ 576 m^2}\right) +H \Big( -B  \frac{q^2}{2}\nonumber\\
    &&-\frac{q^4}{144 m^2}-\alpha^2 \frac{q^2}{36 m} \Big)
       -\sqrt{2} \frac{2}{3} C_{MS}^{S'S' (1)}    +\frac{2}{3} C_{MS}^{S'S' (2)}  \Bigg]\quad,\\
 p_{S'M}^{(u)}&=&\frac{1}{3 \sqrt{2}\alpha^4 } \Bigg[ K \,p_{SS}^{(u)} +L \left( A \left(\frac{3}{2}\alpha^2+\frac{q^2}{8}\right)
  - 5 \alpha^2 \frac{q^2}{36 m^2} -\frac{q^4}{ 144 m^2}\right)+H \Big( -A \frac{q^2}{2}\nonumber\\
     &&+\frac{q^4}{36 m^2}+\alpha^2 \frac{q^2}{9 m} \Big)
       -\sqrt{2} \left( C_{MA}^{S'S' (1)}+\frac{1}{3}  C_{MS}^{S'S' (1)}\right)   +\left( C_{MA}^{S'S' (2)}+\frac{1}{3} C_{MS}^{S'S' (2)}\right)\Bigg]\quad,\nonumber\\
p_{S'M}^{(d)}&=&\frac{1}{ 3 \sqrt{2}\alpha^4 }\Bigg[K\, p_{SS}^{(d)}  +L \left(B \left(\frac{3}{2}\alpha^2+\frac{q^2}{8}\right)+
     5 \alpha^2 \frac{q^2}{144 m^2} +\frac{q^4}{ 576 m^2}\right)\nonumber\\
    && +H \left( -B  \frac{q^2}{2}-\frac{q^4}{144 m^2}-\alpha^2 \frac{q^2}{36 m} \right)
       -\sqrt{2} \frac{2}{3}  C_{MS}^{S'S' (1)}   +\frac{2}{3} C_{MS}^{S'S' (2)} \Bigg]\quad,\\
 p_{MS'}^{(u)}&=&\frac{1}{3\sqrt{2}\alpha^4 } \Bigg[   N\,p_{SS}^{(u)}
       +O\left(A  \left(\frac{3}{2}\alpha^2+\frac{q^2}{8}\right) - 5 \alpha^2 \frac{q^2}{36 m^2} -\frac{q^4}{ 144 m^2}\right)-k_{\lambda}^2 \Big( -A  \frac{q^2}{2}\nonumber\\
       &&+\frac{q^4}{36 m^2}+\alpha^2 \frac{q^2}{9 m} \Big)
       -\sqrt{2} \left(C_{MA}^{S'S' (1)}+\frac{1}{3} C_{MS}^{S'S' (1)}\right)   +\left(C_{MA}^{S'S' (2)}+\frac{1}{3}C_{MS}^{S'S' (2)}\right)\Bigg]\quad,\nonumber\\
 p_{MS'}^{(d)}&=&\frac{1}{3\sqrt{2} \alpha^4 }\Bigg[  N\,p_{SS}^{(d)}  +O \left(B \left(\frac{3}{2}\alpha^2+\frac{q^2}{8}\right)+
     5 \alpha^2 \frac{q^2}{144 m^2} +\frac{q^4}{ 576 m^2}\right) -k_{\lambda}^2 \Big( -B  \frac{q^2}{2}\nonumber\\
    &&-\frac{q^4}{144 m^2}-\alpha^2 \frac{q^2}{36 m} \Big)
       -\sqrt{2} \frac{2}{3} C_{MS}^{S'S' (1)}  +\frac{2}{3}C_{MS}^{S'S' (2)}\Bigg]\quad,\\
  p_{S'M'}^{(u)} &=&-\frac{2}{ 3 \sqrt{18} \,q^2 \alpha^4} \left(C_{MS MA}^{S'M'}+ \frac{1}{3}  C_{MA MS}^{S'M'}\right)\quad, \nonumber\\
   p_{S'M'}^{(d)}   &=&-\frac{2}{3 \sqrt{18} \,q^2  \alpha^4}\frac{2}{3}  \,C_{MA MS}^{S'M'}\quad,\\
   p_{M'S'}^{(u)} &=&-\frac{2}{3\sqrt{18} \,q^2  \alpha^4}\left( C_{MS MA}^{M'S'}+ \frac{1}{3} C_{MA MS}^{M'S'} \right)\quad, \nonumber\\
   p_{M'S'}^{(d)} &=&-\frac{2}{3\sqrt{18} \,q^2  \alpha^4}\frac{2}{3} \,C_{MA MS}^{M'S'}\quad,\\
  p_{MM}^{(u)} &=&\frac{1}{6 \alpha^4 } \Bigg[ S\,  p_{SS}^{(u)} 
       +U \left(A  \left(\frac{3}{2}\alpha^2+\frac{q^2}{8}\right) - 5 \alpha^2 \frac{q^2}{36 m^2} -\frac{q^4}{ 144 m^2}\right)-k_{\lambda}^2 \Big( -A  \frac{q^2}{2}\nonumber\\
       &&+\frac{q^4}{36 m^2}+\alpha^2 \frac{q^2}{9 m} \Big)
       -\sqrt{2} \left(C_{MA}^{S'S' (1)}+\frac{1}{3} C_{MS}^{S'S' (1)}\right)   +\left(C_{MA}^{S'S' (2)}+\frac{1}{3} C_{MS}^{S'S' (2)} \right)\Bigg]\quad,\nonumber\\
    p_{MM}^{(d)} &=&\frac{1}{6 \alpha^4 }\Bigg[  S\,  p_{SS}^{(d)}  +U \left(B \left(\frac{3}{2}\alpha^2+\frac{q^2}{8}\right)+
     5 \alpha^2 \frac{q^2}{144 m^2} +\frac{q^4}{ 576 m^2}\right)-k_{\lambda}^2 \Big( -B  \frac{q^2}{2}\nonumber\\
    && -\frac{q^4}{144 m^2}-\alpha^2 \frac{q^2}{36 m} \Big)
       -\sqrt{2} \frac{2}{3} C_{MS}^{S'S' (1)}   +\frac{2}{3}  C_{MS}^{S'S' (2)} \Bigg]\quad,\\
  p_{M'M'}^{(u)} &=& \frac{2}{3\alpha^4 } \Bigg[
      \left(C_{MS}^{M'M' (1)}+C_{MS}^{M'M' (2)}+C_{MS}^{M'M' (3)}\right)\nonumber\\
   &&+\frac{1}{3} \,\left(C_{MA}^{M'M' (1)}+C_{MA}^{M'M' (2)}+C_{MA}^{M'M' (3)}\right)\Bigg]\quad,\nonumber\\
 p_{M'M'}^{(d)} &=&\frac{2}{3\alpha^4 } \frac{2}{3}  \,\left(C_{MA}^{M'M' (1)}+C_{MA}^{M'M' (2)}+C_{MA}^{M'M' (3)}\right)\quad,\\
p_{MM'}^{(u)}&=&  -\frac{1}{3 \,  q^2\alpha^4 }\,\left(   C_{MA MS}^{MM'}-\frac{1}{3} C_{MS MA}^{MM'} \right)\quad,\nonumber\\
p_{MM'}^{(d)}&=&\frac{1}{3\,q^2 \alpha^4 }\,\frac{2}{3}\, C_{MS MA}^{MM'}\quad,\\
p_{M'M}^{(u)} &=&  -\frac{1}{3\,q^2  \alpha^4}\left(   C_{MA MS}^{M'M}-\frac{1}{3} C_{MS MA}^{M'M} \right)\quad,\nonumber\\
 p_{M'M}^{(d)} &=&\frac{1}{3\,q^2 \alpha^4 }\frac{2}{3} \,C_{MS MA}^{M'M}\quad;
  \eeq           
with 
\beq	
 A&=&\frac{2}{3} + \frac{k_{\lambda}^z}{3 \sqrt 6 m} -  \frac{(q^2-\vec{k}\cdot\vec{q})}{18 m^2}\quad,\nonumber\\
 B&=&-\frac{2}{3} + \frac{k_{\lambda}^z}{3 \sqrt 6 m} +\frac{5}{36 m^2}(q^2-\vec{k}\cdot\vec{q})\quad,\nonumber\\
 C&=&\frac{4}{3}-\frac{k_{\lambda}^z}{3 \sqrt 6 m}-7 \frac{(q^2-\vec{k}\cdot\vec{q})}{36 m^2}\quad,\nonumber\\
  F&=&k_{\lambda}^4+k_{\lambda}^2 q^2-\sqrt{6} k_{\lambda} \vec{q}\cdot\vec{k}_{\lambda}-k_{\lambda}^2 6 \alpha^2-6 q^2
         \alpha^2+3 \alpha^2 \sqrt{6} \vec{q}\cdot\vec{k}_{\lambda}  +9 \alpha^4 \quad,\nonumber\\
 G&=&2 k_{\lambda}^2 +2 q^2-\sqrt{6} \vec{q}\cdot\vec{k}_{\lambda}-6 \alpha^2\quad,\nonumber\\
H&=& k_{\lambda}^2-3 \alpha^2\quad,\nonumber\\
 K&=&-k_{\lambda}^4-k_{\lambda}^2 q^2+\sqrt{6} k_{\lambda}^2
   \vec{q}\cdot\vec{k}_{\lambda}+k_{\lambda}^2 3 \alpha^2+3 q^2 \alpha^2-3 \alpha \sqrt{6} \vec{q}\cdot\vec{k}_{\lambda}\quad,\nonumber\\
 L&=&-q^2+\sqrt{6} \vec{q}\cdot\vec{k}_{\lambda}-3 \alpha^2\quad,\nonumber\\
T&=&1-\frac{\vec{q}\cdot\vec{k}_{\lambda}}{4 \sqrt{6} m^2}\quad,\nonumber\\
T'&=&1-5 \frac{\vec{q}\cdot\vec{k}_{\lambda}}{12 \sqrt{6}m^2}-\frac{k_{\lambda}^z} {\sqrt{6}m}\quad,\nonumber\\
 N&=&-k_{\lambda}^4-2 k_{\lambda}^2 q^2+\sqrt{6} k_{\lambda}^2  \vec{q}\cdot\vec{k}_{\lambda}+k_{\lambda}^2 3 \alpha^2\quad,\nonumber\\
 O&=&2 q^2 -\sqrt{6} \vec{q}\cdot\vec{k}_{\lambda}-3 \alpha^2\quad,\nonumber\\
 S&=&k_{\lambda}^4+ k_{\lambda}^2 q^2-\sqrt{6} k_{\lambda}^2  \vec{q}\cdot\vec{k}_{\lambda}\quad,\nonumber\\
 U&=&- q^2 +\sqrt{6} \vec{q}\cdot\vec{k}_{\lambda}-2k_{\lambda}^2 \quad;
 \eeq
 and with
\beq
C_{MS MA}^{M'S}&=&-q_x+3 q_x \frac{k_{\lambda}^z}{2 \sqrt{6} m} +\frac{k_{\lambda}^x}{2 \sqrt{6} m^2}
   \left(q_x^2-\frac{3}{2} q_y^2\right)+ 5\frac{k_{\lambda}^y q_x q_y}{4 \sqrt{6} m^2 } \quad,\nonumber\\
 C_{MA MS}^{M'S}&=&-q_x+3 q_x \frac{k_{\lambda}^z}{2 \sqrt{6} m} + \frac{k_{\lambda}^x}{2 \sqrt{6} m^2}
   		 \left(q_x^2+\frac{3}{2} q_y^2\right) -\frac{k_{\lambda}^y q_x q_y}{4 \sqrt{6} m^2}\quad,\nonumber\\
   C_{MS MA}^{S'M'} &=&\left(\sqrt{\frac{3}{2}} q^2-\vec{q}\cdot\vec{k}_{\lambda}\right)
   \left(k_{\lambda}^2 -5 \frac{\alpha^2}{2}+\frac{q^2}{8}\right)
    \frac{1}{8 \sqrt{3}}\big( (-2 \sqrt{6}+3 \frac{k_{\lambda}^z}{m}+\frac{\vec{q}\cdot\vec{k}_{\lambda}}{m^2}) q_x\nonumber\\
   &&      -3q_y  \frac{(\vec{k}_{\lambda}\times \vec{q})_z}{2 m^2}\Big)
  -q_x \alpha^2 \frac{k_{\lambda}^z}{4 \sqrt{2}m}
     (k_{\lambda}^2-3 \alpha^2)-q_y (\vec{k}_{\lambda}\times \vec{q})_z \frac{\alpha^2}{16 \sqrt{2}m^2}
   \left( 5 \alpha^2+\frac{q^2}{4}\right)\quad,\nonumber\\
   C_{MA MS}^{S'M'}   &=&\left(\sqrt{\frac{3}{2} }q^2-\vec{q}\cdot\vec{k}_{\lambda}\right)
                    \left(k_{\lambda}^2 -5 \frac{\alpha^2}{2}+\frac{q^2}{8}\right)
                     \frac{1}{8 \sqrt{3}}\Big( (-2 \sqrt{6}+3 \frac{k_{\lambda}^z}{m}+\frac{\vec{q}\cdot\vec{k}_{\lambda}}{m^2}) q_x\nonumber\\
          &&
                   +3q_y  \frac{(\vec{k}_{\lambda}\times \vec{q})_z}{2 m^2}\Big) -q_x \alpha^2 \frac{k_{\lambda}^z}{4 \sqrt{2}m}
     (k_{\lambda}^2-3 \alpha^2)+q_y (\vec{k}_{\lambda}\times \vec{q})_z \frac{\alpha^2}{16 \sqrt{2}m^2}
   \left( 5 \alpha^2+\frac{q^2}{4}\right)\quad,\nonumber\\
    C_{MS MA}^{M'S'}&=&q_x \Bigg[\frac{\vec{q}\cdot\vec{k}_{\lambda}}{2 \sqrt{2}}  \left(-5 \frac{\alpha^2}{2}+13 \frac{q^2}{8}
      +k_{\lambda}^2-\sqrt{6}\vec{q}\cdot\vec{k}_{\lambda}\right)\left(-1+3 \frac{k_{\lambda}^z}{2 \sqrt{6} m}   
                +\frac{\vec{q}\cdot\vec{k}_{\lambda}}{2\sqrt{6} m^2}\right)\nonumber\\
     && +\alpha^2 \frac{k_{\lambda}^z}{4 \sqrt{2} m^2} \left(5 \frac{q^2}{2}+k_{\lambda}^2-\sqrt{6} \vec{q}\cdot\vec{k}_{\lambda}-
       3\alpha^2 \right)\Bigg]\nonumber\\
                 && -q_y\frac{(\vec{k}_{\lambda}\times \vec{q})_z}{ 4\sqrt{2}}\Bigg[ \left(3 \frac{\vec{q}\cdot\vec{k}_{\lambda}}{\sqrt{6}}
                 +\alpha^2\right)\left(13 \frac{q^2}{16} +\frac{k_{\lambda}^2}{2}-\sqrt{6} \frac{\vec{q}\cdot\vec{k}_{\lambda}}{2}\right)
                 -3\frac{5}{4}\alpha^2 \frac{ \vec{q}\cdot\vec{k}_{\lambda}}{\sqrt{6}}-\frac{\alpha^4}{4}\Bigg]\quad,\nonumber\\
     C_{MA MS}^{M'S'}&=&q_x \Bigg[\frac{\vec{q}\cdot\vec{k}_{\lambda}}{2 \sqrt{2}}  \left(-5 \frac{\alpha^2}{2}+13 \frac{q^2}{8}
      +k_{\lambda}^2-\sqrt{6}\vec{q}\cdot\vec{k}_{\lambda}\right)\left(-1+3 \frac{k_{\lambda}^z}{2 \sqrt{6} m}   
                +\frac{\vec{q}\cdot\vec{k}_{\lambda}}{2\sqrt{6} m^2}\right)\nonumber\\
     && +\alpha^2 \frac{k_{\lambda}^z}{4 \sqrt{2} m^2} \left(5 \frac{q^2}{2}+k_{\lambda}^2-\sqrt{6} \vec{q}\cdot\vec{k}_{\lambda}-
       3\alpha^2 \right)\Bigg]\nonumber\\
                 && +q_y\frac{(\vec{k}_{\lambda}\times \vec{q})_z}{ 4\sqrt{2}}\Bigg[ \left(3 \frac{\vec{q}\cdot\vec{k}_{\lambda}}{\sqrt{6}}
                 +\alpha^2\right)\left(13 \frac{q^2}{16} +\frac{k_{\lambda}^2}{2}-\sqrt{6} \frac{\vec{q}\cdot\vec{k}_{\lambda}}{2}\right)
                 -3\frac{5}{4}\alpha^2 \frac{ \vec{q}\cdot\vec{k}_{\lambda}}{\sqrt{6}}-\frac{\alpha^4}{4}\Bigg]\quad,\nonumber\\
     C_{MS MA}^{MM'} &=&q_x \Bigg[\frac{1}{2\sqrt{2}} \left(-1+3 \frac{k_{\lambda}^z}{2 \sqrt{6} m}   
                +\frac{\vec{q}\cdot\vec{k}_{\lambda}}{2\sqrt{6} m^2}\right)\left(\sqrt{\frac{3}{2}} q^2-\vec{q}\cdot\vec{k}_{\lambda}\right)
                \left(-k_{\lambda}^2+\frac{\alpha^2}{2}+ \frac{q^2}{8}\right)\nonumber\\
               && +\alpha^2 k_{\lambda}^2 \frac{k_{\lambda}^z}{4 \sqrt{2} m^2}\Bigg]-q_y (\vec{k}_{\lambda}\times \vec{q})_z \Bigg[ \frac{3}{4\sqrt{6}m^2} \frac{1}{2\sqrt{2}}
               \left(\sqrt{\frac{3}{2}} q^2-\vec{q}\cdot\vec{k}_{\lambda}\right)\left(-k_{\lambda}^2+\frac{\alpha^2}{2}+ \frac{q^2}{8}\right)\nonumber\\
               && 
               +\frac{\alpha^2}{8\sqrt{2} m^2}\left(5\frac{\alpha^2}{2}+\frac{q^2}{8}-k_{\lambda}^2\right)\Bigg]\quad,\nonumber\\
     C_{MA MS}^{MM'}&=&q_x \Bigg[\frac{1}{2\sqrt{2}} \left(-1+3 \frac{k_{\lambda}^z}{2 \sqrt{6} m}   
                +\frac{\vec{q}\cdot\vec{k}_{\lambda}}{2\sqrt{6} m^2}\right)\left(\sqrt{\frac{3}{2}} q^2-\vec{q}\cdot\vec{k}_{\lambda}\right)
                \left(-k_{\lambda}^2+\frac{\alpha^2}{2}+ \frac{q^2}{8}\right)\nonumber\\
                &&+\alpha^2 k_{\lambda}^2 \frac{k_{\lambda}^z}{4 \sqrt{2} m^2}\Bigg]-q_y (\vec{k}_{\lambda}\times \vec{q})_z \Bigg[ \frac{3}{4\sqrt{6}m^2} \frac{1}{2\sqrt{2}}
               \left(\sqrt{\frac{3}{2}} q^2-\vec{q}\cdot\vec{k}_{\lambda}\right)\left(-k_{\lambda}^2+\frac{\alpha^2}{2}+ \frac{q^2}{8}\right)\nonumber\\
                &&
               +\frac{\alpha^2}{8\sqrt{2} m^2}\left(5\frac{\alpha^2}{2}+\frac{q^2}{8}-k_{\lambda}^2\right)\Bigg]\quad,\nonumber\\
 C_{MS MA}^{M'M}&=&q_x \Bigg[\frac{\vec{q}\cdot\vec{k}_{\lambda}}{2\sqrt{2}} \left(-1+3 \frac{k_{\lambda}^z}{2 \sqrt{6} m}   
                +\frac{\vec{q}\cdot\vec{k}_{\lambda}}{2\sqrt{6} m^2}\right)
                \left(\sqrt{6} \vec{q}\cdot\vec{k}_{\lambda} -k_{\lambda}^2+2 \alpha^2+11 \frac{q^2}{8}\right)\nonumber\\
               && -\alpha^2 \frac{k_{\lambda}^z}{4 \sqrt{2} m^2} \left(\sqrt{6} \vec{q}\cdot\vec{k}_{\lambda} 
                -k_{\lambda}^2-3\frac{q^2}{2}\right)\Bigg]\nonumber\\
            && -q_y (\vec{k}_{\lambda}\times \vec{q})_z \Bigg[3\frac{\vec{q}\cdot\vec{k}_{\lambda}}{8\sqrt{12} m^2}\left(\sqrt{6} \vec{q}\cdot\vec{k}_{\lambda} 
            -k_{\lambda}^2+2 \alpha^2+11 \frac{q^2}{8}\right)\nonumber\\
            &&+\frac{\alpha^2}{8\sqrt{2}m^2}  \left(\sqrt{6} \vec{q}\cdot\vec{k}_{\lambda} -k_{\lambda}^2+5\frac{q^2}{2}-11 \frac{q^2}{8}\right)\Bigg]
            \quad,\nonumber\\
C_{MA MS}^{M'M}&=&q_x \Bigg[\frac{\vec{q}\cdot\vec{k}_{\lambda}}{2\sqrt{2}} \left(-1+3 \frac{k_{\lambda}^z}{2 \sqrt{6} m}   
                +\frac{\vec{q}\cdot\vec{k}_{\lambda}}{2\sqrt{6} m^2}\right)\left(\sqrt{6} \vec{q}\cdot\vec{k}_{\lambda}
                 -k_{\lambda}^2+2 \alpha^2+11 \frac{q^2}{8}\right)\nonumber\\
                &&-\alpha^2 \frac{k_{\lambda}^z}{4 \sqrt{2} m^2} \left(\sqrt{6} \vec{q}\cdot\vec{k}_{\lambda} 
                -k_{\lambda}^2-3\frac{q^2}{2}\right)\Bigg]\nonumber\\
            && +q_y (\vec{k}_{\lambda}\times \vec{q})_z \Bigg[3\frac{\vec{q}\cdot\vec{k}_{\lambda}}{8\sqrt{12} m^2}\left(\sqrt{6} \vec{q}\cdot\vec{k}_{\lambda} 
            -k_{\lambda}^2+2 \alpha^2+11 \frac{q^2}{8}\right)\nonumber\\
            &&+\frac{\alpha^2}{8\sqrt{2}m^2}  \left(\sqrt{6} \vec{q}\cdot\vec{k}_{\lambda} -k_{\lambda}^2+5\frac{q^2}{2}-11 \frac{q^2}{8}\right)\Bigg]
            \quad,
 \eeq

 \beq 
 C_{MA}^{S'S' (1)} &=&q^2 \Big(-5 \frac{\alpha^4}{16 \sqrt{2} m^2}
      +\alpha^2\left(5 \frac{T}{4 \sqrt{2}}-\frac{q^2}{8 \sqrt{2} m^2}\right)+T \frac{q^2}{16 \sqrt{2}}
        -\frac{q^4}{256 \sqrt{2} m^2}\Big)\quad,\nonumber\\
  C_{MA}^{S'S' (2)} &=&q^2 \Big(-70 \frac{\alpha^4}{128 m^2} +\alpha^2 \left(5 \frac{T}{8}-\frac{7 q^2}{128 m^2}\right)
              +T \frac{q^2}{64}-\frac{q^4}{128 8 m^2}\Big)\quad,\nonumber\\
 C_{MS}^{S'S' (1)}  &=&-q^2 \Big(-5 \frac{\alpha^4}{48 \sqrt{2} m^2}
   +\alpha^2\left(5 \frac{T'}{4 \sqrt{2}}-\frac{q^2}{24 \sqrt{2} m^2}\right)+T' \frac{q^2}{16 \sqrt{2}}
        -\frac{q^4}{768 \sqrt{2} m^2}\Big)\quad,\nonumber\\
C_{MS}^{S'S' (2)} &=&-q^2 \Big(-70 \frac{\alpha^4}{384 m^2} +\alpha^2 \left(5 \frac{T'}{8}-\frac{7 q^2}{384 m^2}\right)
              +T' \frac{q^2}{64}-\frac{q^4}{384 8 m^2}\Big)\quad,\nonumber\\
 C_{MA}^{M'M' (1)}  &=&\frac{1}{4}\left(\sqrt{\frac{3}{2}} q^2-\vec{q}\cdot\vec{k}_{\lambda}\right)\left(
     \alpha^2 \frac{k_{\lambda}^z}{m}+\vec{q}\cdot\vec{k}_{\lambda} T-\alpha^2 \frac{ \vec{q}\cdot\vec{k}_{\lambda}}{4 m^2}
       -q^2 \frac{ \vec{q}\cdot\vec{k}_{\lambda}}{16 m^2}\right)\quad,\nonumber\\
    C_{MS}^{M'M' (1)} &=&-\frac{1}{4}\left(\sqrt{\frac{3}{2}} q^2-\vec{q}\cdot\vec{k}_{\lambda}\right)\left(
     \frac{2}{3}\alpha^2 \frac{k_{\lambda}^z}{m}+\vec{q}\cdot\vec{k}_{\lambda} T'-\alpha^2 \frac{ \vec{q}\cdot\vec{k}_{\lambda}}{12 m^2}
       -q^2 \frac{ \vec{q}\cdot\vec{k}_{\lambda}}{48 m^2}\right)\quad,\nonumber\\
 C_{MA}^{M'M' (2)}&=&-\sqrt{\frac{3}{2}} \left(T \vec{q}\cdot\vec{k}_{\lambda}\left(\frac{\alpha^2}{2}+ \frac{q^2}{8}\right)+\alpha^2 q^2 \frac{k_{\lambda}^z}{8m}
  - 3 \vec{q}\cdot\vec{k}_{\lambda}\frac{ \alpha^2 q^2}{32 m^2}-q^4 \frac{\vec{q}\cdot\vec{k}_{\lambda}}{128  m^2}\right)\quad,\nonumber\\
  C_{MS}^{M'M' (2)}&=&\sqrt{\frac{3}{2}} \left(T' \vec{q}\cdot\vec{k}_{\lambda}\left(\frac{\alpha^2}{2}+ \frac{q^2}{8}\right)+\alpha^2 q^2\frac{2}{3}
    \frac{k_{\lambda}^z}{8m}
  -  \vec{q}\cdot\vec{k}_{\lambda}\frac{ \alpha^2 q^2}{32 m^2}-q^4 \frac{\vec{q}\cdot\vec{k}_{\lambda}}{384  m^2}\right)\quad,\nonumber\\
 C_{MA}^{M'M' (3)}&=&T \frac{\alpha^2 k_{\lambda}^2}{2} +\frac{T }{8}(\vec{q}\cdot\vec{k}_{\lambda})^2
     -\alpha^2 \left(\vec{q}\cdot\vec{k}_{\lambda} \frac{k_{\lambda}^z}{4 m}+q^2 \frac{ k_{\lambda}^2}{32 m^2}+
  	\frac{(\vec{q}\cdot\vec{k}_{\lambda})^2}{16m^2} \right)\nonumber\\
	&&-\frac{q^2 (\vec{q}\cdot\vec{k}_{\lambda})^2}{128 m^2}\quad,\nonumber\\
 C_{MS}^{M'M' (3)}&=&	-\Bigg [T' \frac{\alpha^2 k_{\lambda}^2}{2} +\frac{T' }{8}(\vec{q}\cdot\vec{k}_{\lambda})^2
     -\alpha^2 \left(\frac{2}{3}\vec{q}\cdot\vec{k}_{\lambda} \frac{k_{\lambda}^z}{4 m}+q^2 \frac{ k_{\lambda}^2}{96 m^2}+
  	\frac{(\vec{q}\cdot\vec{k}_{\lambda})^2}{48 m^2} \right)\nonumber\\
	&&-\frac{q^2 (\vec{q}\cdot\vec{k}_{\lambda})^2}{384 m^2}\Bigg]\quad.
\eeq

\acknowledgments
\vskip 2mm
We are grateful to S. Noguera for fruitful discussions.
S.S. thanks the Department of Theoretical Physics of the Valencia
University for warm hospitality; 
A.C. 
and V.V. thank the Department of Physics of the Perugia
University for warm hospitality.  We would like to thank 
A. Metz and P. Schweitzer for sending us the parameterization of the data.  
One of us (S.S.)  thanks
A. Drago and G. Salm\`e for useful conversations.
V.V. and A. C.  would like to thank
Igor O. Cherednikov for fruitful discussions. 
This work is supported in part by the INFN-CICYT agreement,
by the Generalitat Valenciana under the contract
AINV06/118; by the Sixth Framework Program of the
European Commision under the Contract No. 506078 (I3 Hadron Physics);
by the MEC (Spain) under the Contract FPA 2007-65748-C02-01 and
the grants AP2005-5331 and PR2007-0048.

\newpage

\section*{Figure Captions}

\vspace{1em}\noindent
{\bf Fig. 1}:
The contributions to the Sivers
function in the present approach. The graph has been
drawn using JaxoDraw \cite{Binosi:2003yf}.

\vspace{1em}\noindent
{\bf Fig. 2}: In the upper (lower) panel,
the quantity $\rho_Q(\vec b)$, Eq. (\ref{rhob}),
is shown for the $u$ ($d$) flavor.

\vspace{1em}\noindent
{\bf Fig. 3}:
The quantity $f_{1T}^{\perp (1)q }(x) $, Eq. (\ref{momf}),
for the $u$ flavor.
The dashed curve is the result of the present approach
at the hadronic scale $\mu_0^2$, Eq. (\ref{result}).
The full curve represents the evolved distribution after
standard NLO evolution (see text).
The patterned area represents the $1 - \sigma$ range of the best fit 
of the HERMES data
proposed in Ref. \cite{coll3}.

\vspace{1em}\noindent
{\bf Fig. 4}: The same as in Fig. 3, but for the $d$ flavor.

\newpage

\begin{figure}[ht]
\includegraphics{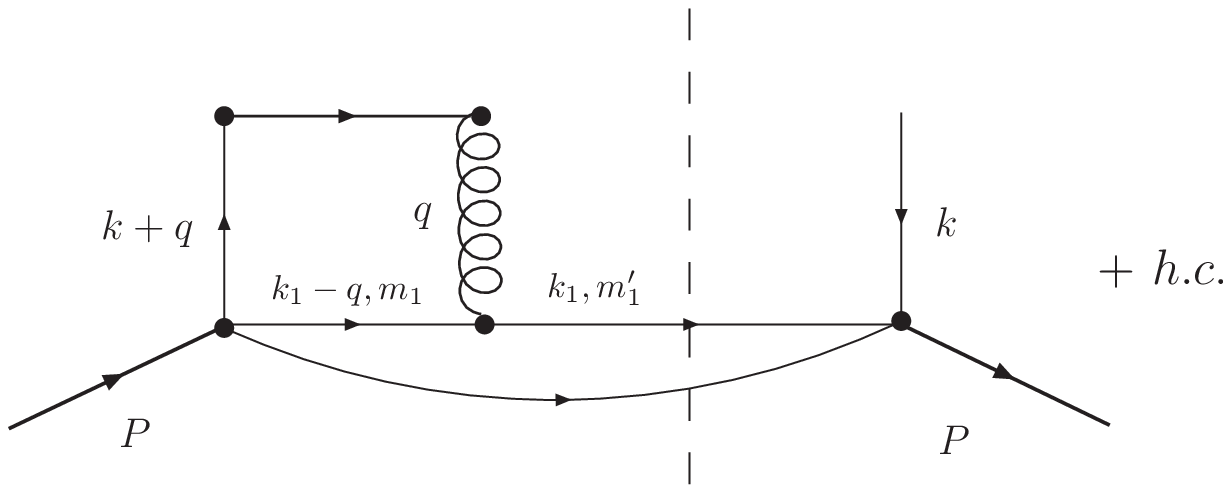}
\vspace{12.0cm}
\caption{}
\end{figure}

\newpage

\begin{figure}[ht]
\includegraphics{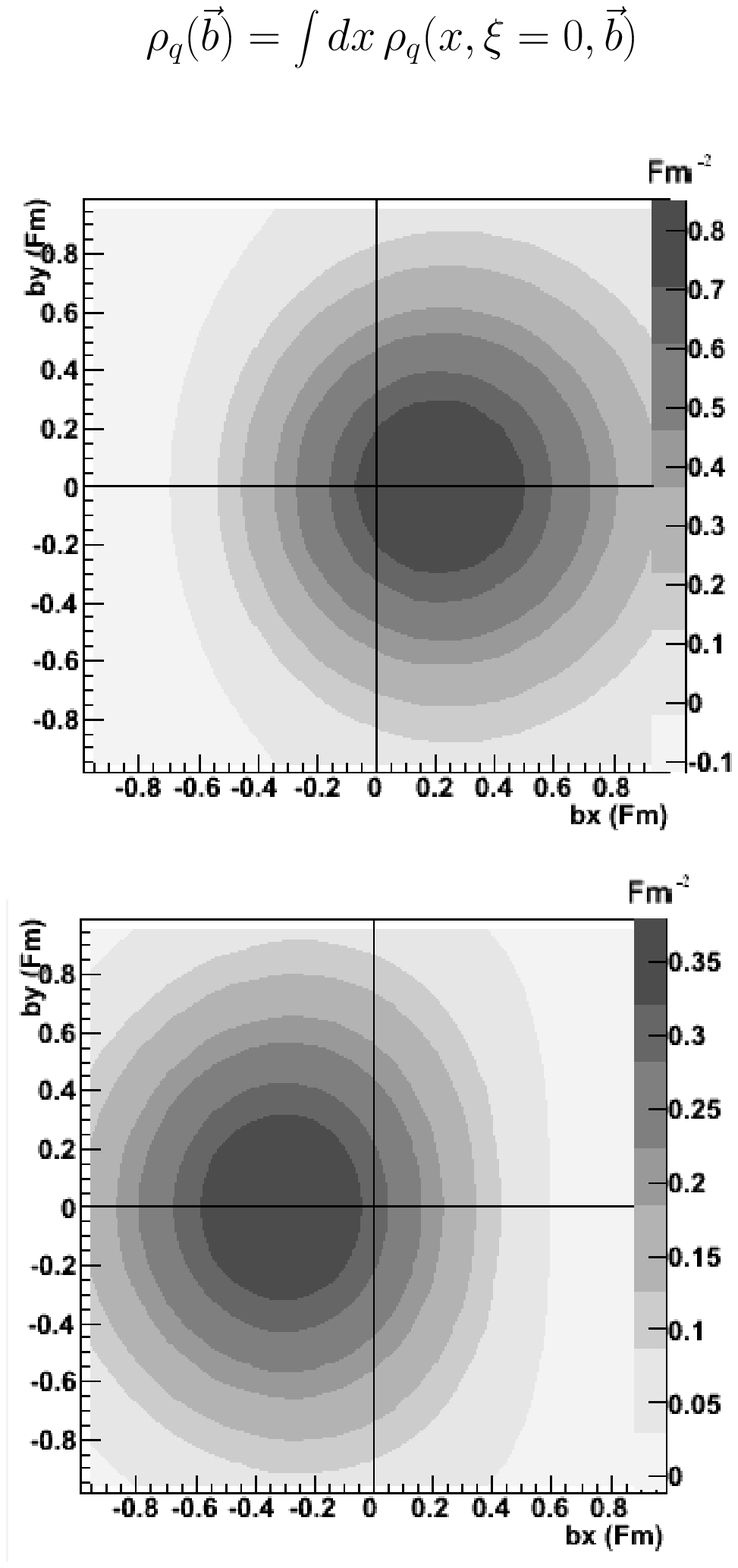}
\vspace{12.0cm}
\caption{}
\label{filippo}
\end{figure}

\newpage

\begin{figure}[h]
% \vspace{10cm}
%\special{psfile=siv_u_coll2.eps hoffset=0 voffset=-700 hscale=80 vscale=80}
\includegraphics{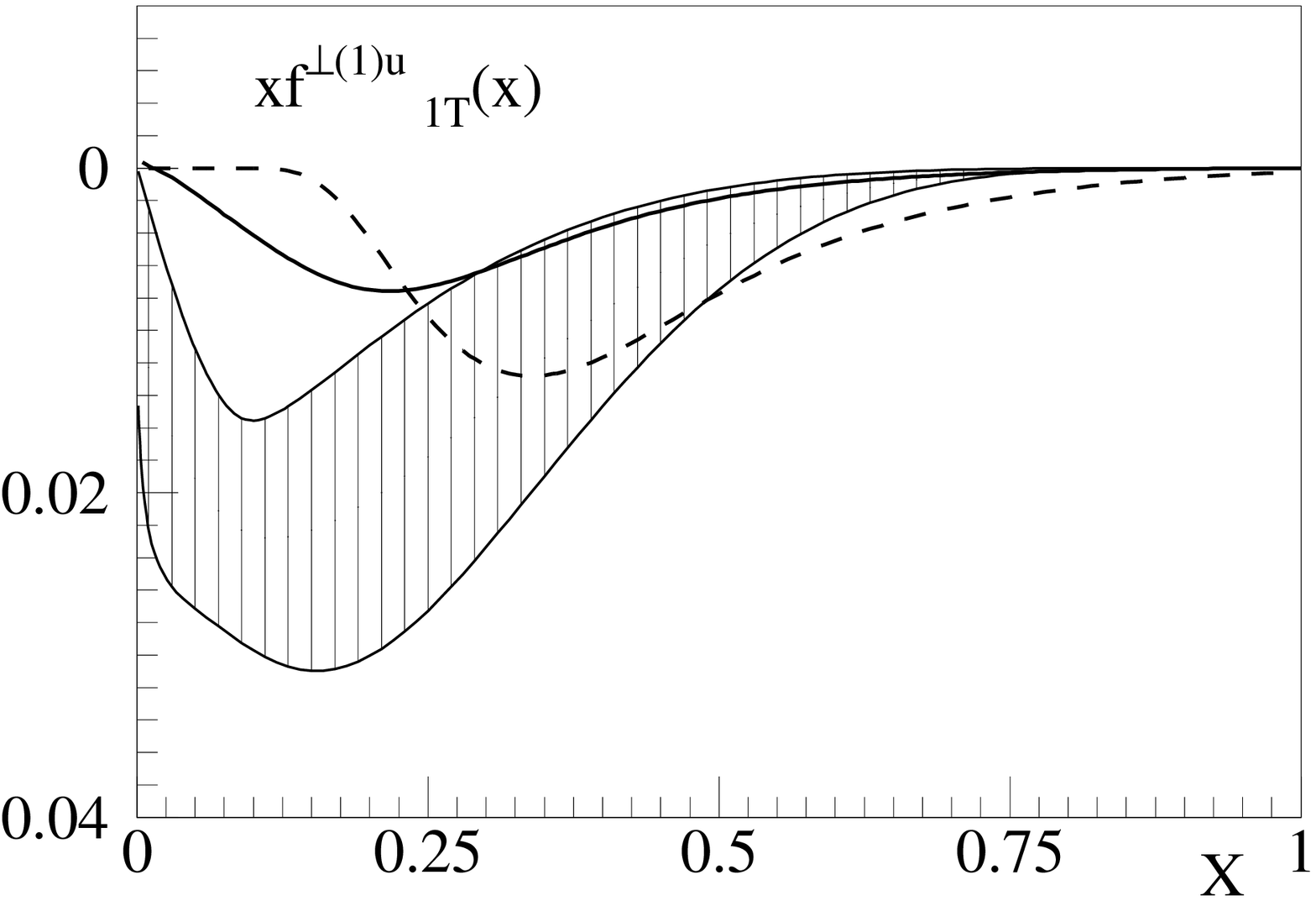}
\vspace{12.0cm}
\caption{}
\end{figure}

\newpage

\begin{figure}[h]
% \vspace{10cm}
%\special{psfile=siv_d_coll2.eps hoffset=0 voffset=-700 hscale=80 vscale=80}
\includegraphics{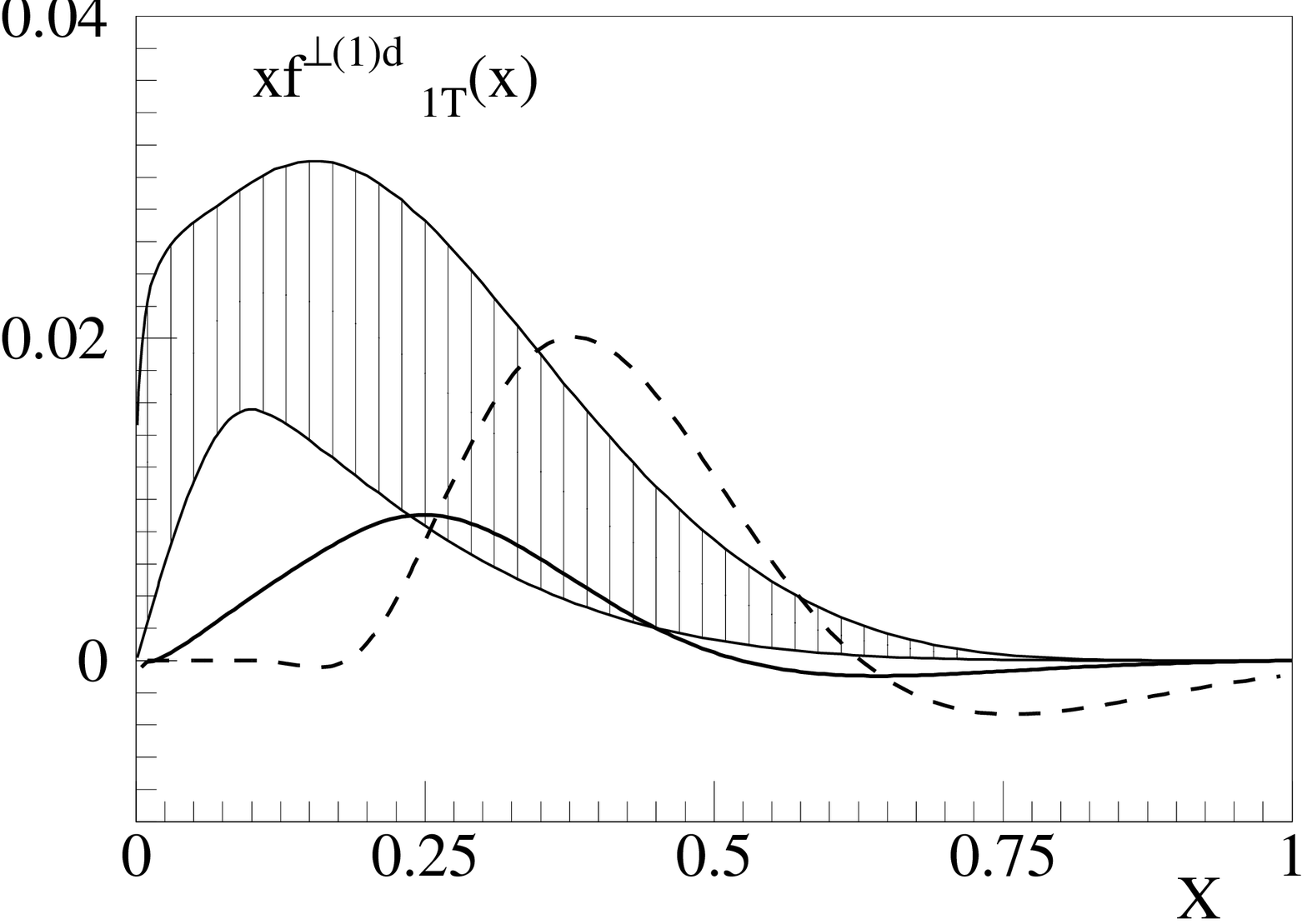}
\vspace{12.0cm}
\caption{}
\end{figure}

\end{document}